\documentclass[aps, prb, superscriptaddress, reprint, twocolumn, amsmath, amssymb, notitlepage]{revtex4-2}

\usepackage{amsmath,bm}
\usepackage{dcolumn}
\usepackage{mathrsfs}
\usepackage{graphicx}% Include figure files
\usepackage{dcolumn}% Align table columns on decimal point
\usepackage{bm}% bold math
\usepackage[capitalise]{cleveref}
\usepackage{siunitx}
\usepackage{amsfonts}%
\usepackage{soul,color}
\begin{document}
\title{Theory of Linewidth-Narrowing in Fano Lasers}% Force line breaks with \\
%%\thanks{A footnote to the article title}%
\author{Yi Yu}
%%\altaffiliation[Also at ]{Physics Department, XYZ University.}%Lines break automatically or can be forced with \\
 \email{yiyu@dtu.dk}
\author{Aref Rasoulzadeh Zali}%
\author{Jesper M{\o}rk}%
 \email{jesm@dtu.dk}
\affiliation{DTU Electro, Technical University of Denmark, 2800 Kgs. Lyngby, Denmark}%
\affiliation{NanoPhoton - Center for Nanophotonics, Technical University of Denmark, Ørsteds Plads 345A, DK-2800 Kgs. Lyngby, Denmark}
% \date{\today}% It is always \today, today,
%              %  but any date may be explicitly specified

\begin{abstract}
We present a general theory for the coherence of Fano lasers based on a bound state in the continuum. We find that such lasers enable orders of magnitude reduction of the quantum-limited linewidth, and by introducing mirror symmetry breaking, the linewidth can be further reduced, In contrast to ordinary macroscopic lasers, though, the linewidth may re-broaden due to optical nonlinearities enhanced by the strong light localization. This leads to the identification of optimal material systems. We also show that the coherence of this new type of microscopic laser can be understood intuitively using a simple, effective potential model. Based on this model, we examine the laser stability and deduce the dependence of the laser linewidth on the general Fano lineshape. Our model facilitates the incorporation of other degrees of design freedom and can be applied to a general class of lasers with strongly dispersive mirrors.  
% \begin{description}
% \item[Usage]
% Secondary publications and information retrieval purposes.
% \item[Structure]
% You may use the \texttt{description} environment to structure your abstract;
% use the optional argument of the \verb+\item+ command to give the category of each item. 
% \end{description}
\end{abstract}

%\keywords{Suggested keywords}%Use showkeys class option if keyword
                              %display desired
\maketitle

%\tableofcontents

% \section{\label{sec:level1}First-level heading:\protect\\ The line
% break was forced \lowercase{via} \textbackslash\textbackslash}

\section{Introduction}
Realizing ultra-coherent nanoscale lasers is important for numerous applications, including on-chip communications \cite{Sun2015}, programmable photonic integrated circuits \cite{Bogaerts2020}, bio/chemical sensing \cite{Ge2013}, and quantum and neuromorphic computing \cite{Carolan2015,Shen2017}. However, since the quantum-limited laser linewidth scales inversely with the number of photons in the cavity, the small mode volume of nanolasers poses a fundamental challenge in realizing highly coherent microscopic lasers \cite{Bjork1992,Ellis2011,Kim2012,Nezhad2010,Azzam2020}. Indeed, ultra-small metallic lasers show linewidths of tens of GHz \cite{Nezhad2010}, while state-of-the-art semiconductor lasers \cite{Septon2019}, albeit with millimeter lengths, can achieve linewidths less than 1 kHz \cite{Santis2018}. Nevertheless, a linewidth of a few MHz was recently demonstrated in a microlaser by taking advantage of the unusual properties of a bound state in the continuum (BIC) \cite{Yu2021}. One of the laser mirrors thus uses a Fano resonance to implement the highly dispersive characteristic of a BIC. This dispersion reduces the laser linewidth by several orders of magnitude without introducing a large cavity. The concept of such a Fano laser is generic \cite{Limonov2017}, and in this paper, we present a general theory for the quantum-limited linewidth of this new type of laser. Compared to Ref.~\cite{Yu2021}, in which the response function of the Fano mirror is a symmetric Lorentzian corresponding to the specific case of a Fano-shape parameter $q$=0 \cite{Fano1961}, we here extend the theory to account for general Fano lineshapes \cite{Miroshnichenko2010,Suh2004,Bekele2019}. The new theory can be cast in a form, where the instantaneous laser frequency behaves analogously to a particle moving in a potential and exposed to a randomly fluctuating force, representing quantum noise. Such analogy was initially developed for external cavity lasers \cite{MORK1992E,Mark1992}, which today represent the most important type of narrow-linewidth laser, but is here extended to the generic case of a Fano laser. We predict that by introducing mirror symmetry breaking, further linewidth reduction by a factor of four is possible. Surprisingly, we find that optical nonlinearities may give rise to a fundamental limitation to the linewidth. 

Our theory not only offers a simple and intuitive explanation of the physics of linewidth reduction in Fano lasers, but also highlights the qualitative differences between Fano lasers and external cavity lasers and may inspire future innovations. 

\section{Structure and theoretical model}
\begin{figure}
\centering
\setlength{\abovecaptionskip}{3pt} % Chosen fairly arbitrarily
\includegraphics[width=1\linewidth]{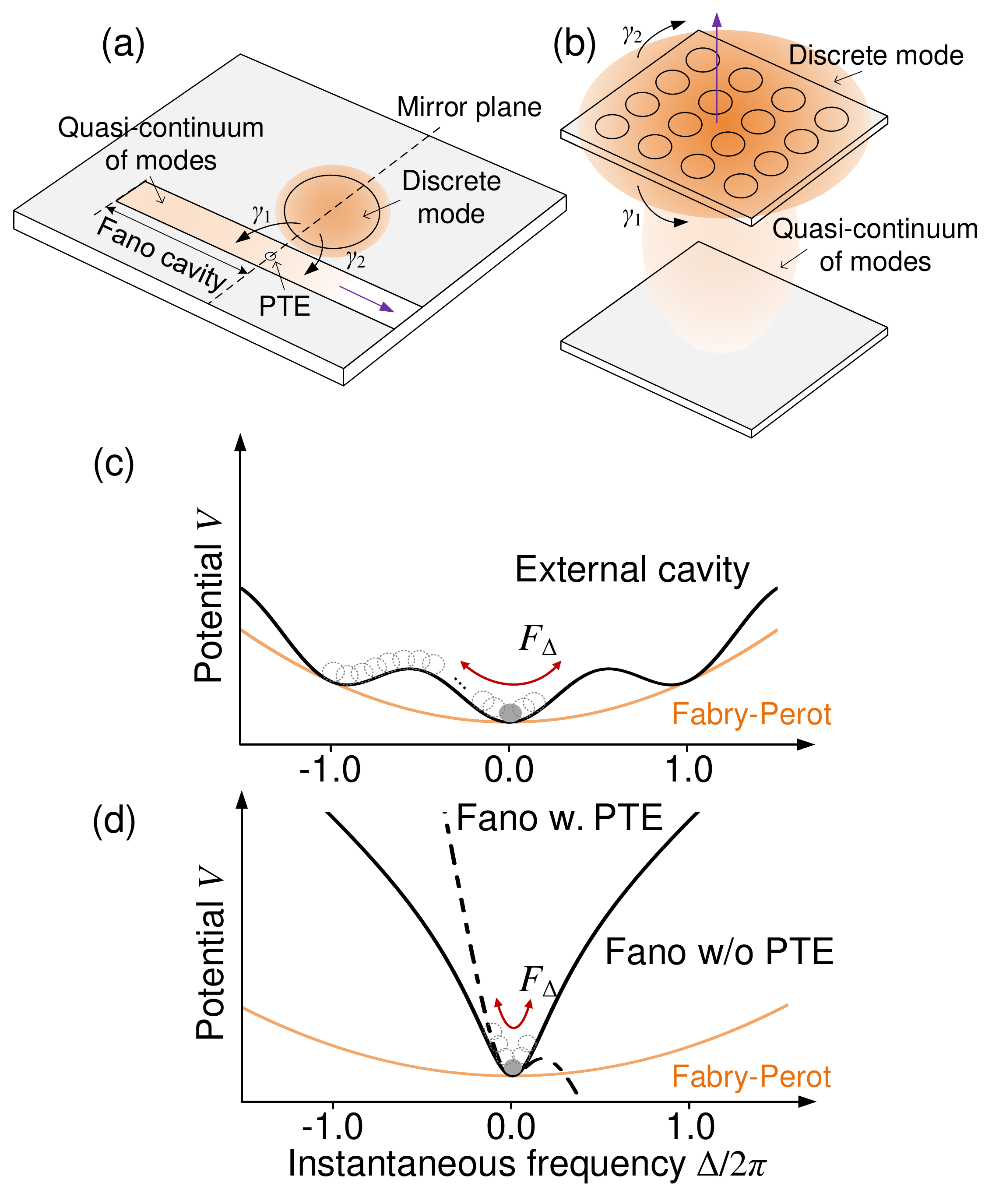}
\caption{\label{fig:v1}(a) Fano laser realized in an in-plane geometry. A partially transmitting element (PTE), with a field reflectivity of $r_B$, is placed in the region hosting a continuum of modes. The gain material of the laser is placed in the continuum region, while the region hosting the discrete mode is passive. (b) Fano laser realized in a vertical configuration, where a photonic crystal slab is used as the narrowband Fano mirror, and serves as the combination of the PTE and discrete state in the photonic crystal band diagram. The orange shadings illustrate the lasing field, with the darker area corresponding to stronger field intensity. (c, d) Phase potential $V$ as a function of instantaneous frequency $\Delta $ for (c) a conventional external cavity laser and (d) the Fano laser. The laser experiences phase noise represented by a Langevin force ${{F}_{\Delta }}$. The black curves represent (c) the external cavity laser and (d) the Fano laser, and the orange curves represent the Fabry-Perot laser counterpart. For the Fano laser, the solid (dashed) black curve corresponds to the case without (with) a PTE. Here, ${{\gamma }_{in}}/{{\gamma }_{D}}$ takes the values of 5000 and 78 for (c) and (d), respectively.}
\end{figure}

The Fano laser (Figs.~\ref{fig:v1}(a,~b)) is constructed by coupling a discrete mode with a (quasi-) continuum of modes, which can be implemented in various configurations. For example, it can be realized in an in-plane design (see Fig.~\ref{fig:v1}(a)) by a photonic crystal membrane with a line-defect semi-open waveguide (WG) and a right mirror realized by a nanocavity adjacent to the WG \cite{Mork2014,Yu2017,Rasmussen2017,Mork2019,Rasmussen2019}. A field propagating to the right in the WG can take two paths. One path follows the WG, while the other comprises tunneling through the nanocavity. These two paths interfere destructively around the nanocavity resonance wavelength, leading to a high reflectivity within a narrow bandwidth, which we refer to as a Fano mirror. A virtual cavity (Fano cavity) can now be formed between the left mirror and the Fano mirror and has the characteristics of a BIC. Thus, the mode only forms if the phase condition is fulfilled (the roundtrip phase change must be an integer multiple of 2$\pi$) at the resonance of the nanocavity, where the Fano mirror reflectivity is high. This leads to a sensitive geometry-dependency of the quality factor ($Q$-factor) of the cavity mode, which is a hallmark of a BIC \cite{Rybin2017}. Such a Fano laser has been realized using buried-heterostructure technology \cite{Yu2021}, where the active material is embedded only in the WG, however, the laser is of a general nature and can also be realized in other platforms using various technologies, such as monolithic integration \cite{Wang2015,Mayer2019}, heterogeneous integration \cite{Tran2019}, or hybrid integration \cite{Li2017nn,Birowosuto2014}. In addition to the in-plane configuration, the Fano laser can also be realized in a vertical configuration (see Fig.~\ref{fig:v1}(b)), e.g., using a dielectric slab as a broadband mirror and a photonic crystal slab as the narrowband Fano mirror \cite{Fan2002}. As our interest is on-chip applications, we here focus on the in-plane configuration. 

The dynamics of the Fano laser can be described by coupled-mode equations \cite{Haus} combined with conventional rate equations \cite{Mork2014,Yu2021}. By assuming that the left mirror has unity reflectivity, we have
\begin{align} 
\frac{d}{dt}{{A}^{+}}\left( t \right)&=\left( \left( 1-j\alpha  \right){{G}_{N}}\Delta N\left( t \right)-{{\gamma }_{in}} \right){{A}^{+}}\left( t \right)\nonumber \\ 
&+{{\gamma }_{in}}{{A}^{-}}\left( t \right)/{{r}_{R}}\left( {{\omega }_{r}} \right)+{{F}_{{{A}^{+}}}}\left( t \right),\label{eq:s1} 
\end{align}
\begin{align} 
\frac{d}{dt}N\left( t \right)&=R-{{\gamma }_{N}}N\left( t \right)\nonumber \\
&-{{G}_{N}}\left( N\left( t \right)-{{N}_{0}} \right)I\left( t \right)/{{V}_{a}}+{{F}_{N}}\left( t \right),\label{eq:s2} 
\end{align}
where ${{A}^{+}}\left( t \right)$ (${{A}^{-}}\left( t \right)$) is the slowly varying complex amplitude of the forward (backward) propagating field, $\alpha $ is Henry’s factor \cite{Henry1982}, ${{G}_{N}}$ is the modal gain factor, $\omega $ (${{\omega }_{r}}$) is the lasing frequency (reference frequency, e.g., the steady-state lasing frequency excluding quantum noise), $N\left( t \right)$ is the carrier density in the Fano cavity, ${{N}_{0}}$ is the carrier density at transparency, $\Delta N\left( t \right)=N\left( t \right)-{{N}_{s}}$ is the carrier density deviation from its steady-state value ${{N}_{s}}$, and ${{\gamma }_{N}}$ is the carrier decay rate. Furthermore, $R$, ${{V}_{a}}$, and ${{\gamma }_{in}}$=1/$\tau_{in}$ denote the pumping rate, the active volume, and the round-trip rate of the Fano cavity, respectively. The number of photons stored within the Fano cavity is related to the complex field, $I(t)={{\varsigma }_{s}}\left( {{\omega }_{r}} \right){{\left| {{A}^{+}}\left( t \right) \right|}^{2}}$, with the conversion factor obtained from the steady-state solution \cite{Tromborg1987}, ${{\varsigma }_{s}}\left( {{\omega }_{r}} \right)=\left( 1-{{\left| {{r}_{R}}\left( {{\omega }_{r}} \right) \right|}^{2}} \right)/\left( 2\hbar {{\omega }_{r}}{{\gamma }_{in}}\ln \left\{ 1/\left| {{r}_{R}}\left( {{\omega }_{r}} \right) \right| \right\} \right)$. The backward field is obtained from the nanocavity field, ${{A}_{c}}\left( t \right)$, which in turn is driven by the forward-propagating field
\begin{align} 
&{{A}^{-}}\left( t \right)={{r}_{B}}{{A}^{+}}\left( t\right)+\sqrt{2{{\gamma}_{1}}}{{A}_{c}}\left( t \right),\label{eq:s3}
\end{align}
\begin{align} 
\frac{d}{dt}{{A}_{c}}\left( t \right)&=\left( -j\left( \delta_{0} +{{\delta }_{NL}}\left( t \right) \right)-{{\gamma }_{t}} \right){{A}_{c}}\left( t \right)\nonumber \\ 
&+\sqrt{2{{\gamma }_{1}}}{{e}^{2j{{\theta }_{1}}}}{{A}^{+}}\left( t \right)+{{F}_{{{A}_{c}}}}\left( t \right).\label{eq:s4}
\end{align}
The nanocavity field is normalized such that ${{\left| {{A}_{c}}\left( t \right) \right|}^{2}}$ is the energy stored in the nanocavity. The parameter ${{r}_{B}}$ (${{t}_{B}}=\sqrt{1-r_{B}^{2}}$) is the reflectivity (transmissivity) of the partially transmitting element (PTE in Fig.~\ref{fig:v1}(a)), which will be discussed later. In addition, $\delta_{0} ={{\omega }_{0}}-{{\omega }_{r}}$ is the detuning between the nanocavity resonance ${{\omega }_{0}}$ and ${{\omega }_{r}}$, ${{\delta }_{NL}}\left( t \right)$ is the complex change of the nanocavity resonance due to optical nonlinearities leading to intensity-dependent detuning and loss \cite{Yu2013}. Furthermore, ${{\gamma }_{v}}$ is the nanocavity intrinsic decay rate, and ${{\gamma }_{1}}$ (${{\gamma }_{2}}$) is the nanocavity coupling rate to the left (right) side of the WG (Fig.~\ref{fig:v1}(a)), with ${{\gamma }_{c}}={{\gamma }_{1}}+{{\gamma }_{2}}$ and ${{\gamma }_{t}}={{\gamma }_{v}}+{{\gamma }_{c}}$. These (field amplitude) decay rates are related to the nanocavity intrinsic, coupling, and total $Q$-factors as $Q_v=\omega_0/(2\gamma_v)$, $Q_c=\omega_0/(2\gamma_c)$, and $Q_t=\omega_0/(2\gamma_t)$. In Eqs.~(\ref{eq:s1})-(\ref{eq:s4}), $F_{A^+}\left( t \right)$, $F_{A_c}\left( t \right)$, and $F_N\left( t \right)$ are the Langevin noise terms of the fields and carrier density. The coefficient ${{e}^{2j{{\theta }_{1}}}}$, depending on the coupling phase, can be derived by exploiting energy conservation and time-reversal symmetry \cite{Suh2004,Yu2015}, leading to $\cos \left( 2{{\theta }_{1}} \right)={{\gamma }_{2}}t_{B}^{2}/\left( 2{{\gamma }_{1}}{{r}_{B}} \right)-t_{B}^{2}/\left( 2{{r}_{B}} \right)-{{r}_{B}}$ and $\sin \left( 2{{\theta }_{1}} \right)=-P{{t}_{B}}\sqrt{4{{\gamma }_{1}}{{\gamma }_{2}}-t_{B}^{2}\gamma _{c}^{2}}/\left( 2{{\gamma }_{1}}{{r}_{B}} \right)$. It should be noted that the transmission through a linear Fano mirror follows the general expression for a Fano lineshape characterized by the shape parameter $q$ \cite{Yoon2013}: $T\left( \delta_{0}  \right)=t_{B}^{2}{{\left( q+\delta_{0} /{{\gamma }_{t}} \right)}^{2}}/\left( 1+{{\left( \delta_{0} /{{\gamma }_{t}} \right)}^{2}} \right)$. Here, $q=-\tan \left( \vartheta  \right)$ with $\vartheta =-P{{\cos }^{-1}}\left( -{{t}_{B}}{{\gamma }_{c}}/\left( 2\sqrt{{{\gamma }_{1}}{{\gamma }_{2}}} \right) \right)$. 

\section{Potential picture of the Fano laser frequency}
For linewidth narrowing, we exploit that the BIC has a large fraction of its electromagnetic energy stored in the nanocavity. Now, if the gain material of the laser is excluded from the nanocavity, the field stored in the nanocavity is not exposed to spontaneous emission and acts as a restoring force that counteracts the random phase changes induced by spontaneous emission in the spatial regions containing gain. In this case, by defining the instantaneous laser frequency $\Delta =\left( {{\omega }_{r}}-\omega  \right){{\tau }_{D}}$, where ${{\tau }_{D}}$ is the time delay caused by the right mirror, and separating the amplitude and phase of the laser fields, one can derive from Eqs.~(\ref{eq:s1})-(\ref{eq:s4}) (see Appendix A) that the laser frequency behaves analogously to a particle moving with strong friction in a potential $V$ and exposed to random kicks representing spontaneous emission, cf. Figs.~\ref{fig:v1}(c,~d). For a Fano laser, we have
\begin{equation}
{{V}}={{V}_{FP}}+{{V}_{FM}}, \label{eq:a1}
\end{equation}
where ${{V}_{FP}}={{\Delta }^{2}}/\left( 2{{\tau }_{D}} \right)$ is the potential of an equivalent Fabry-Perot (FP) laser with cavity length equal to the Fano cavity, and frequency-independent mirror reflectivities. The other term in Eq.~(\ref{eq:a1}) is ${{V}_{FM}}={{\gamma }_{in}}\ln \left( 1+{{\Delta }^{2}} \right)/2$ and originates from the frequency-dependency of the Fano mirror. The laser linewidth ($\Delta v$) is represented by the frequency spread resulting from the random excursions around the potential minimum and is inversely proportional to the square of the potential curvature, i.e., $1/\Delta v\propto {{\left( {{d}^{2}}V/d{{\Delta }^{2}} \right)}^{2}}$. 

This is similar to the case where an FP laser is coupled to a (large) passive external cavity (Fig.~\ref{fig:v1}(c)), where $V={{V}_{FP}}+{{V}_{EX}}$ and ${{V}_{EX}}=-\kappa {{\gamma }_{in}}\cos \left( {{\theta }_{0}}+\Delta  \right)$ \cite{Mark1992} with $\kappa $ being the feedback fraction. As seen, an external cavity introduces a cosine function adding to the original parabolic function of the FP laser \cite{Mark1992}. As a consequence, several potential valleys appear, corresponding to different external cavity modes that the laser jumps between on time scales determined by the height of the potential barriers \cite{MORK1992E}. In contrast, the Fano laser has a single minimum corresponding to the single BIC.  Furthermore, compared to the cosine function of ${{V}_{EX}}$, ${{V}_{FM}}$ is a logarithmic function and does not depend on $\kappa $, which is usually much smaller than unity \cite{Mark1992,Li1989}. Therefore, the Fano laser configuration is much more effective in narrowing the laser linewidth, and furthermore accomplishes it without introducing a large external cavity or a large secondary resonator \cite{Tran2019}, which would severely increase the footprint of the laser.

If the symmetry of the Fano resonance is changed, which can be achieved by adding a PTE to change the amplitude and phase of the continuum path \cite{Heuck2013}, additional linewidth narrowing can be achieved but at the prize of sacrificing the monostability of the laser (asymmetric potential in Fig.~\ref{fig:v1}(d)). The detailed picture is that, compared to the case without a PTE, the presence of the PTE lowers $V$ on one side. This means that the laser tends to have multiple solutions, rather than being monostable, e.g., the laser can jump to another solution as the right potential barrier is passed. This new solution corresponds to $2\omega nL/c+\arg \left\{ {{r}_{R}}\left( \omega  \right) \right\}=2m\pi $ where $n$ and $L$ are the refractive index and length of the Fano cavity, respectively, and $m$ is an integer different from the value corresponding to the original Fano mode. When the laser oscillation frequency moves away from the nanocavity resonance, $\left| {{r}_{R}}\left( \omega  \right) \right|$ decreases toward 0 ($r_B$) when the PTE is absent (present), corresponding to a diverging (finite) laser threshold. Therefore, the PTE increases the chance of mode hopping into another longitudinal mode of the Fano cavity \cite{Mork2019}. However, it should be noted that the multiple longitudinal modes appearing in a Fano laser with a PTE cannot be captured here, since our model is derived based on the expansion around a single longitudinal mode \cite{Yu2017}. 

The field reflectivity of the Fano mirror ${{r}_{R}}\left( \omega  \right)$, with the presence of a PTE, has the general form \cite{Yu2015}
\begin{align}
&{{r}_{R}}\left( \omega  \right)=\left| {{r}_{R}}\left( \omega  \right) \right|{{e}^{j{{\phi }_{R}}\left( \omega  \right)}}\nonumber \\ 
&=\frac{jr_{B}^{2}\delta +\left( {{\gamma }_{2}}-{{\gamma }_{1}} \right)+r_{B}^{2}{{\gamma }_{v}}-jP{{t}_{B}}\sqrt{4{{\gamma }_{1}}{{\gamma }_{2}}-t_{B}^{2}\gamma _{c}^{2}}}{{{r}_{B}}\left( j\delta +{{\gamma }_{t}} \right)}. \label{eq:a2} 
\end{align}
Here, $\delta ={{\delta }_{0}}+{{\delta }_{NL}}$, with ${{\delta }_{0}}={{\omega }_{0}}-\omega $. The parity of the nanocavity mode with respect to the mirror plane, cf. Fig.~\ref{fig:v1}(a), is accounted for by the coefficient, $P$, with $P$=1 (-1) corresponding to an even (odd) mode of the nanocavity and leading to a red (blue) parity for an asymmetric Fano resonance with ${{r}_{B}}\ne 0$ \cite{Yu2014,Yu2015}. Compared to the case without the PTE, where the decay ratio ${{R}_{12}}={{\gamma }_{1}}/{{\gamma }_{2}}=1$, the PTE can break the nanocavity mirror symmetry (${{\gamma }_{1}}/{{\gamma }_{2}}\ne 1$), e.g., by incorporating a hole in the WG away from the mirror plane \cite{Osterkryger2016}. This can enlarge the phase slope of ${{r}_{R}}\left( \omega  \right)$ without sacrificing the mirror reflectivity $\left| {{r}_{R}}\left( \omega  \right) \right|$ at the nanocavity resonance (see Fig.~4 in Appendix A), indicating that the laser linewidth can be further reduced by the inclusion of a PTE.

From the laser model, the lasing frequency $\omega $ fulfills the relation ${{\omega }_{r}}=\omega -\alpha {{\gamma }_{in}}+\alpha \operatorname{Re}\left\{ K{{r}_{R}}\left( \omega  \right) \right\}+\operatorname{Im}\left\{ K{{r}_{R}}\left( \omega  \right) \right\}$ with $K={{\gamma }_{in}}/{{r}_{R}}\left( {{\omega }_{r}} \right)$. Based on the perturbation approach \cite{Li1989}, one can derive the Fano laser linewidth above the threshold
\begin{align}
& \Delta {{v}_{FL}}\left( {{\omega }_{r}} \right)=\Delta {{v}_{FP}}\left( {{\omega }_{r}} \right)/{{\eta }^{2}},\nonumber \\ 
& \eta =1+{{\gamma }_{in}}{{\left. \left( \alpha \frac{1}{\left| {{r}_{R}}\left( \omega  \right) \right|}\frac{\partial }{\partial \omega }\left| {{r}_{R}}\left( \omega  \right) \right|+\frac{\partial }{\partial \omega }{{\phi }_{R}}\left( \omega  \right) \right) \right|}_{\omega ={{\omega }_{r}}}}, \label{eq:a3}
\end{align}
where $\Delta {{v}_{FP}}\left( {{\omega }_{r}} \right)=\frac{\left( 1+{{\alpha }^{2}} \right){{\left( {{\gamma }_{i}}+{{\gamma }_{in}}\ln \left\{ 1/\left| {{r}_{R}}\left( {{\omega }_{r}} \right) \right| \right\} \right)}^{2}}{{C}_{sp}}}{R-{{\gamma }_{N}}{{N}_{th}}\left( {{\omega }_{r}} \right)}$ is the linewidth of the equivalent FP laser and $\eta $ is the linewidth reduction factor. Here, ${{\gamma }_{i}}$ is the loss rate of the active WG, and ${{C}_{sp}}={{n}_{sp}}/\left( \pi {{V}_{a}} \right)$ is a coefficient depending on the population inversion factor, ${{n}_{sp}}$. The threshold carrier density is given by ${{N}_{th}}\left( {{\omega }_{r}} \right)={{N}_{0}}+\left( {{\gamma }_{i}}+{{\gamma }_{in}}\ln \left\{ 1/\left| {{r}_{R}}\left( {{\omega }_{r}} \right) \right| \right\} \right)/{{G}_{N}}$. As seen from Eq.~(\ref{eq:a3}), the $\alpha $-parameter can also contribute to linewidth narrowing, provided it has the same sign as $\partial \left| {{r}_{R}}\left( \omega  \right) \right|/\partial \omega $, which leads to negative feedback between the laser frequency change and the refractive index change induced by the change of free carrier density \cite{Kazarinov1987}. Eq.~(\ref{eq:a3}) is in accordance with the previous result \cite{Yariv1995} and can be related to the effective $Q$-factor of the Fano laser. 

To evaluate the effective $Q$-factor (${Q}_{FL}$) of the Fano laser, we consider at the electromagnetic energy stored in the entire Fano laser, $E_{FL}$, which consists of the energy stored in the Fano cavity, ${{E}_{F}}$, and the energy stored in the nanocavity, ${{E}_{nc}}$. By neglecting the waveguide dispersion, the stored energies are proportional to the group delay of each part \cite{Winful2003}, ${{\tau }_{in}}$ and ${{\tau }_{D}}$, as
\begin{align}
  & {{E}_{FL}}={{E}_{F}}+{{E}_{nc}}={{E}_{FP}}\frac{{{\tau }_{in}}+{{\tau }_{nc}}}{{{\tau }_{in}}} \nonumber \\ 
 & ={{E}_{FP}}{{\left. \left( 1+\frac{\partial }{\partial \omega }\arg \left( {{r}_{R}}\left( \omega  \right) \right)/\frac{\partial }{\partial \omega }\left( \frac{2L\omega n}{c} \right) \right) \right|}_{\omega ={{\omega }_{r}}}}. \nonumber \\ \label{eq:s117}
\end{align}
Here, ${{E}_{FP}}$ is the energy stored in a FP cavity with a dimension equivalent to the Fano cavity (FP counterpart). ${{\left. \frac{\partial }{\partial \omega }\left( \frac{2L\omega n}{c} \right) \right|}_{\omega ={{\omega }_{r}}}}={{\tau }_{in}}$. The time averaged power dissipation $P_{s}$ is caused by the propagation loss in the Fano cavity and the power leakage from the right mirror. The value of this quantity is the same for both the Fano laser and the FP counterpart, leading to
\begin{align}
{{Q}_{FL}}={{\omega }_{r}}\frac{{{E}_{FL}}}{P_{s}}={{Q}_{FP}}\left( {{\left. 1+{{\gamma }_{in}}\frac{\partial }{\partial \omega }\arg \left( {{r}_{R}}\left( \omega  \right) \right) \right|}_{\omega ={{\omega }_{r}}}} \right). \nonumber
\end{align}
Here, ${Q}_{FP}$ is the $Q$-factor of the FP cavity laser. Eq.~(\ref{eq:s117}) slightly deviates from the one derived in Ref.~\cite{Yu2021}. We think this is because we here take into account the energy (usually negligible) stored in the WG on the right side of the left reference plane of the Fano mirror, which is also part of the lasing mode.

In the following, we investigate how the response function of the Fano resonance affects the laser linewidth. The simulation parameters, unless specified, are kept fixed: ${{Q}_{c}}$=1000, ${{\omega }_{0}}$=$2\pi c/{{\lambda }_{0}}$ with ${{\lambda }_{0}}$=1550 nm, $\alpha $=3, ${{\gamma }_{i}}$=4.7$\times {{10}^{10}}\ $s$^{-1}$, ${{\gamma }_{in}}$=9.5×10$^{12}$ s$^{-1}$, ${{\gamma }_{N}}$=4.2$\times$10$^8$ s$^{-1}$, ${{G}_{N}}$=5.2$\times $10$^{-13}$ m$^3$/s, $N_0$=0.4$\times $10$^{24}$ m$^{-3}$, $C_{sp}$=1.1$\times $10$^{20}$ m$^{-3}$. These values are in accordance with the results in Ref.~\cite{Yu2017,Yu2021}. For simplicity, we assume negligible intrinsic losses of the isolated nanocavity, corresponding to ${{Q}_{v}}$→$\infty$ (a finite value, $Q_v\ge$10$^5$, does not change the picture). In addition, we chose $P$=1, i.e., the Fano resonance has blue parity, which leads to narrower laser linewidth for the common case in semiconductors, where $\alpha $ is positive \cite{Tran2019,Jin2021}.

\section{Fano laser linewidth in the linear case}
We first focus on the linear case where ${{\delta }_{NL}}=0$. As seen from Eq.~(\ref{eq:a3}) and the expression for ${{N}_{th}}\left( {{\omega }_{r}} \right)$, the Fano laser linewidth can, for fixed pumping, be reduced by either reducing the linewidth of the solitary FP laser by increasing the Fano mirror reflectivity ${{\left| {{r}_{R}}\left( {{\omega }_{r}} \right) \right|}^{2}}$ (to lower the laser threshold), or by increasing the linewidth reduction factor by increasing the (absolute) value of the (normalized) amplitude differential ${{L}_{1}}\left( \delta  \right)=2\pi \alpha {{\left| {{r}_{R}}\left( \omega  \right) \right|}^{-1}}{{\left. \left( \frac{\partial }{\partial \omega }\left| {{r}_{R}}\left( \omega  \right) \right| \right) \right|}_{\omega ={{\omega }_{r}}}}$, or the phase differential ${{L}_{2}}\left( \delta  \right)=2\pi {{\left. \left( \frac{\partial }{\partial \omega }{{\phi }_{R}}\left( \omega  \right) \right) \right|}_{\omega ={{\omega }_{r}}}}$. All three quantities can be controlled by $\delta $ and ${{R}_{12}}$. In general, for a given ${{R}_{12}}$, large values for $\left| {{L}_{1}}\left( \delta  \right) \right|$ ($\left| {{L}_{2}}\left( \delta  \right) \right|$) are found on the low (high) reflectivity side of the Fano mirror (see Fig.~4 in Appendix A). Examples of the Fano laser linewidth variation with laser operation frequency are shown in Fig.~\ref{fig:v2}(a). A pump power of $R={{10}^{35}}$m$^{-3}$s$^{-1}$ is applied, which is $\sim$37 dB above the lowest laser threshold obtained for ${{R}_{12}}=1$. Such a high pump power is beyond (about 10 dB higher than) our current experimental possibilities, and is not a requirement for achieving narrow linewidth but is chosen to ensure that lasing can occur even around the minimum reflectivity coefficient of the Fano mirror. Sub-MHz linewidths can already be achieved at $R< 2\times 10^{33}$m$^{-3}$s$^{-1}$, even with a relatively low nanocavity $Q$-factor of 1000. 

For $\alpha \ne 0$, the laser linewidth exhibits two local minima. The right minimum, with a sharp spectral feature, corresponds to the reflectivity minimum, and the linewidth narrowing here is due mainly to the amplitude-phase coupling, where a large ${{L}_{1}}\left( \delta  \right)$ (because of a large value of $\partial \left| {{r}_{R}}\left( \omega  \right) \right|/\partial \omega $ accompanied by a small value of $\left| {{r}_{R}}\left( \delta  \right) \right|$) provides negative feedback, suppressing laser frequency fluctuations. The left local linewidth minimum, with a broader spectral feature, occurs close to the reflectivity maximum of the Fano mirror and is due to a combination of low threshold (high $\left| {{r}_{R}}\left( \omega  \right) \right|$) and prolonged photon storage in the passive nanocavity (large ${{L}_{2}}\left( \delta  \right)$). For this case, the influence of ${{L}_{1}}\left( \delta  \right)$ is negligible. 

The laser linewidth also assumes a local maximum, which is blue or red detuned with respect to the linewidth minimum, depending on the sign of $\alpha $. Such a maximum corresponds to $\eta=0$, where $L_1\left( \delta  \right)=-\left( 2\pi /{{\gamma }_{in}}+{{L}_{2}}\left( \delta  \right) \right)$. 
\onecolumngrid
\begin{center}
\setlength{\abovecaptionskip}{3pt} % Chosen fairly arbitrarily
\begin{figure}[h]
\includegraphics[scale=0.199]{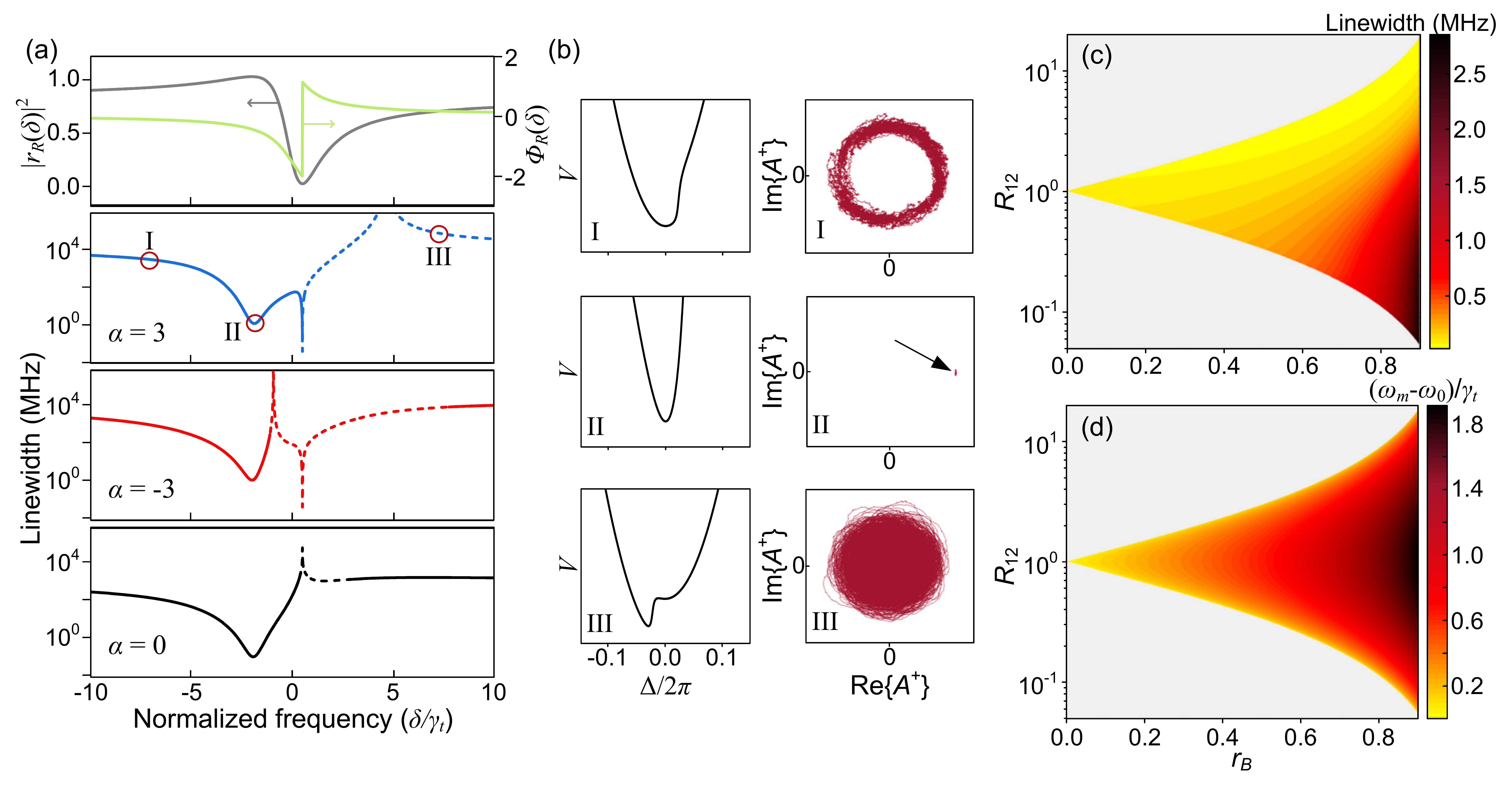}
\caption{\label{fig:v2}(a) Calculated Fano laser linewidths $\Delta {{v}_{FL}}$ (black, red, and blue curves) versus (normalized) lasing frequency for different values of the $\alpha $-parameter. The frequency dependence of the Fano mirror reflectivity and phase are also shown (upper gray and green curves). The solid (dashed) parts of the curves represent stable (unstable) solutions. Here, $r_{B}^{2}$=0.8 and $R_{12}$=1. (b) Phase potential $V$ as a function of instantaneous frequency $\Delta $ (left) and the resulting temporal trajectory of the complex field $A^{+}$ of the Fano laser (right), corresponding to the spectral positions I, II, and III for $\alpha $=3 in (a). The trajectory of $A^{+}$ in case II will evolve into a "fuzzy" circle as the simulation time is prolonged. The trajectory of $A^{+}$ in case III corresponds to self-pulsations. (c) The minimum Fano laser linewidth and (d) the corresponding lasing frequency $\omega_{m}$ as a function of $r_B$ and $R_{12}$ for $\alpha $=3. The grey parameter regions are not accessible. In all the cases, a high pump power of $R={{10}^{35}}$m$^{-3}$s$^{-1}$ is assumed.}
\end{figure}
\end{center}
\twocolumngrid
\noindent 

Here, we will focus on the good-cavity case, where the $Q$-factor is high, so that the Peterman-factor \cite{Petermann1979} is close to unity and the dispersion of the gain material can be neglected; these effects will become important in the bad-cavity case and lead to correction factors of Eq.~(\ref{eq:a3}) \cite{Pick2015}. It should also be pointed out that Eq.~(\ref{eq:a3}) only is valid for stable solutions. The laser stability can be checked by a linear stability analysis or direct numerical simulations including Langevin terms (see Appendix C), and can for certain kinds of instability (saddle-point instabilities) also be inferred from the laser potential \cite{Mark1992}. Large fluctuations of the Langevin force may cause the laser to switch from one local potential minimum to another minimum or cause switching to another solution, which is not described by the same effective potential, similar to the case of external cavity lasers \cite{morkptl1990}. Considering that the right local minimum corresponds to a very large laser threshold and can lead to laser instabilities, we in the following focus only on the left minimum.  

Figs.~\ref{fig:v2}(c,~d) show, in color, the minimum linewidth and the corresponding lasing frequency ${{\omega }_{m}}$ versus $r_B$ and $R_{12}$. The grey regions cannot be accessed since $R_{12}$ is bounded by ${{R}_{\min }}\le {{R}_{12}}\le {{R}_{\max }}$, where ${{R}_{\max }}=\left( 1+{{r}_{B}} \right)/\left( 1-{{r}_{B}} \right)$ and ${{R}_{\min }}=1/{{R}_{\max }}$ \cite{Wang2013}. As seen, the smallest linewidth always occurs at ${{\lambda }_{m}}\approx {{\lambda }_{0}}$ with ${{R}_{12}}={{R}_{\max }}$, corresponding to the upper edge of the contour of Fig.~\ref{fig:v2}(c) (and $q$ = 0 for the general Fano formula). This reflects that the main factors contributing to the laser linewidth are $\left| {{r}_{R}}\left( \delta  \right) \right|$ and ${{L}_{2}}\left( \delta  \right)$, which are large at $\delta =0$ in the case of ${{R}_{12}}={{R}_{\max }}$ (see Fig.~4 in Appendix A).   

Next, we focus on highly asymmetric structures, i.e., the upper edge of the contour of Fig.~\ref{fig:v2}(c). Based on Eq.~(\ref{eq:a3}), utilizing $\delta \ll{{\gamma }_{c}}$ and ${{R}_{12}}={{R}_{\max }}$, a simple expression for the Fano laser linewidth can be derived
\begin{equation}
\Delta {{v}_{FL}}\left( {{\omega }_{m}} \right)={{\left( \frac{1}{1+\left( 1+{{r}_{B}} \right)\left( {{\gamma }_{in}}/{{\gamma }_{t}} \right)} \right)}^{2}}\Delta {{v}_{FP}},\label{eq:a4}
\end{equation}
where
\begin{equation}
{{\omega }_{m}}={{\omega }_{0}}\left( 1-\frac{\alpha {{\gamma }_{i}}}{t_{B}^{2}{{\omega }_{0}}\left( 1+\left( 1+{{r}_{B}} \right)\left( {{\gamma }_{in}}/{{\gamma }_{t}} \right) \right)} \right)\approx {{\omega }_{0}}. \label{eq:a5}
\end{equation}
Eq.~(\ref{eq:a4}) shows that compared to the corresponding FP laser, with a linewidth of $\Delta {{v}_{FP}}\approx \left( 1+{{\alpha }^{2}} \right){{C}_{sp}}\gamma _{i}^{2}/R$ under high-pumping assumption, where $R-{{\gamma }_{N}}{{N}_{th}}\left( {{\omega }_{r}} \right)\approx R$, the Fano mirror can significantly improve the laser coherence. Such an improvement becomes more pronounced as ${{\gamma }_{in}}$ (${{\gamma }_{t}}$) increases (decreases), e.g., by reducing the Fano cavity size or improving the nanocavity $Q$-factor. Compared to conventional external cavity lasers \cite{Li1989,Liang2015}, which are based on weak feedback with a linewidth of $\Delta {{v}_{ext}}={{\left( 1/\left( 1+\kappa \left( {{\gamma }_{in}}/{{\gamma }_{t}} \right) \right) \right)}^{2}}\Delta {{v}_{FP}}$, where $\kappa \ll1$, the Fano laser intrinsically operates in the regime of strong feedback and thereby enables a much larger linewidth reduction without significantly increasing the size of the laser. At the same time, the Fano laser does not experience any modal doublets, in contrast to ordinary strong injection-locking or feedback systems \cite{Jin2021}. 

For ultrasmall lasers, where ${{\gamma }_{in}}/{{\gamma }_{T}}\gg1$, Eq.~(\ref{eq:a4}) shows that by breaking the mirror symmetry, $\Delta {{v}_{FL}}\left( {{\omega }_{m}} \right)$ can be further reduced by a (maximum) factor of four, compared to the ordinary symmetric case (${{R}_{12}}=1$). This is because a higher $r_B$ enables a larger ${{R}_{\max }}$ and thus a larger ${{\gamma }_{1}}$ for a fixed $Q_c$, leading to a larger $\left| {{A}_{c}}\right|$ \cite{Yu2015o} (compared to the case of $r_B$=0, the nanocavity field gets doubled under the condition of $r_B$=1 and ${{R}_{12}}={{R}_{\max }}$). This can also be understood in another way: to achieve an effective Fano destructive interference at the output of the WG, the decay of the nanocavity field to the right side, $\left| \sqrt{2{{\gamma }_{2}}}{{A}_{c}} \right|$, should be balanced by the field $\left| j{{t}_{B}}{{A}^{+}} \right|$ transmitted directly through the WG. Therefore, a smaller ${{\gamma }_{2}}$, a larger $\left| {{A}_{c}} \right|$, or a stronger field localization in the passive nanocavity region. This enhances the laser’s composite $Q$-factor. This is consistent with Figs.~4(a,~b), in which ${{\left| {{r}_{R}}\left( {{\omega }_{0}} \right) \right|}^{2}}$ approaches unity with the frequency slope doubled as ${{R}_{12}}\to {{R}_{\max }}$. 

In the case of a high PTE reflectivity, the Fano laser may appear to be equivalent to a system of two coupled cavities. However, in contrast to the case of two coupled cavities, the Fano laser mode still bears the characteristics of a BIC, even for $r_B$=1, where the PTE completely blocks the right end of the WG. This can be concluded by analyzing the Fano mode as a superposition of two coupled modes (see Appendix B). 

The linewidth expressed by Eq.~(\ref{eq:a4}) is identical to the result derived using the Langevin approach (see Appendix C), where stochastic Langevin noises are introduced for the Fano cavity field ${{A}^{+}}\left( t \right)$. The absolute output power is, of course, also important. It can be shown that the external quantum efficiency is much higher for the cross-port \cite{Mork2019} and the left mirror rather than the port involving transmission through the Fano mirror (see Appendix D).

\section{Fano laser linewidth in the nonlinear case}
The theory and results presented so far assumed the nanocavity to have a linear response, i.e., ${{\delta }_{NL}}\left( t \right)$=0. However, the spatially localized field in the nanocavity of the Fano laser can induce a large power-dependent change in the laser output by spectrally shifting or changing the amplitude of ${{r}_{R}}\left( {{\omega }_{r}} \right)$ through optical nonlinearities \cite{Yu2013}. We incorporate nonlinear absorption and index changes by taking 
\begin{align}
{{\delta }_{NL}}\left( t \right)=\left( {{K}_{K}}-j{{K}_{T}} \right){{\left| {{A}_{c}}\left( t \right) \right|}^{2}}+\left( {{K}_{D}}-j{{K}_{A}} \right){{N}_{c}}\left( t \right). \label{eq:a10}
\end{align}
Here, ${{K}_{K}}{{\left| {{A}_{c}}\left( t \right) \right|}^{2}}$ and ${{K}_{D}}{{N}_{c}}\left( t \right)$ account for the nanocavity resonance shifts due to Kerr and free-carrier effects, with ${{K}_{K}}$ being the Kerr coefficient, and ${{K}_{D}}$ accounting for free-carrier dispersion and bandfilling \cite{Yu2013}. The terms ${{K}_{T}}{{\left| {{A}_{c}}\left( t \right) \right|}^{2}}$ and ${{K}_{A}}{{N}_{c}}\left( t \right)$ account for absorption, with ${{K}_{T}}$ being the two-photon absorption (TPA) coefficient and ${{K}_{A}}$ the free carrier absorption coefficient. Furthermore, ${{N}_{c}}\left( t \right)$ is the mode-averaged free carrier density generated by TPA in the nanocavity, governed by 
\begin{align}
\partial {{N}_{c}}\left( t \right)/\partial t=-{{\gamma }_{nc}}{{N}_{c}}\left( t \right)+{{G}_{T}}{{\left| {{A}_{c}}\left( t \right) \right|}^{4}}, \label{eq:a11}
\end{align}
where ${{\gamma }_{nc}}$ is the effective carrier decay rate in the nanocavity and ${{G}_{T}}$ is the free carrier generation coefficient due to TPA. The coefficients ${{K}_{K}}$, ${{K}_{T}}$, ${{G}_{T}}$ and ${{\gamma }_{nc}}$ depend on the nonlinear optical mode volumes of the nanocavity \cite{Yu2013}. We compare photonic crystal L7 nanocavities made of InP, Si, and SiN working at $\sim$1.55 µm, and choose ${{\gamma }_{in}}$=1.9$\times$10$^{12}$ s$^{-1}$ and ${{\gamma }_{nc}}=2\times {{10}^{10}}$s$^{-1}$. We calculate the linewidth including nonlinearities also using the potential approach (see Eqs.~(\ref{eq:s112}) and (\ref{eq:s122}) in Appendix A), with parameters for the three material systems considered (see TABLE I in Appendix C), which agrees well with the conventional Langevin approach (see Appendix C).
\begin{figure}[tbp!]
\centering
\setlength{\abovecaptionskip}{3pt} % Chosen fairly arbitrarily
\includegraphics[width=1.0\linewidth]{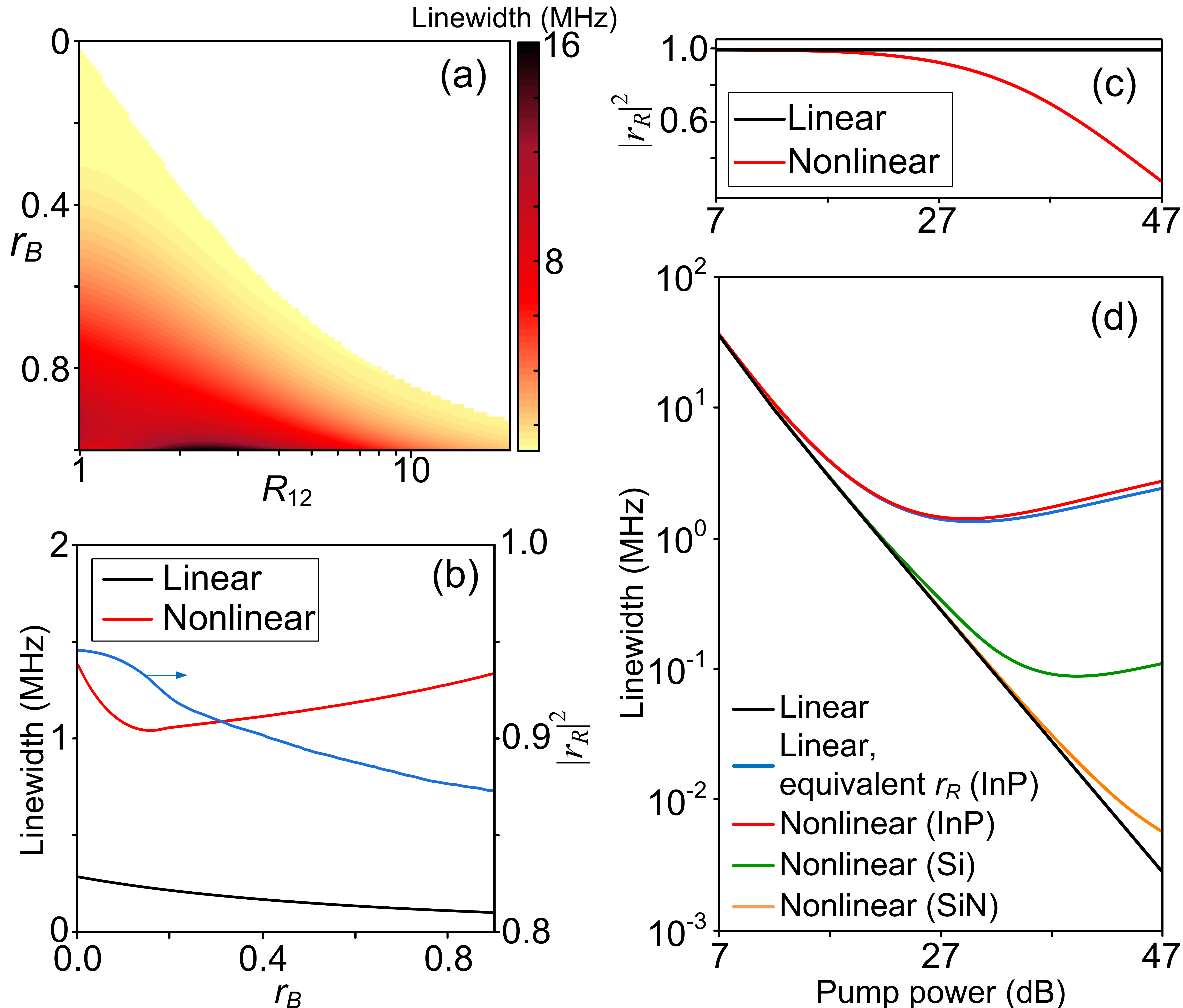}
\caption{\label{fig:v3}(a) Fano laser linewidth $\Delta {{v}_{FL}}\left( {{\omega }_{0}} \right)$ (color scale) as a function of $r_B$ and ${{R}_{12}}$ when nanocavity nonlinearities are accounted for. Here, the nanocavity is made of InP, and a relatively high pump power of $R$=10$^{35}$ m$^{-3}$s$^{-1}$ is chosen to better illustrate the nonlinear effects. (b) The minimum linewidth extracted from Fig.~\ref{fig:v3}(a) (red curve) and Fig.~\ref{fig:v2}(c) (black curve) as a function of $r_B$. The Fano mirror reflectivity ${{\left| {{r}_{R}}\left( {{\omega }_{0}} \right) \right|}^{2}}$ (blue curve) corresponding to the red curve is also plotted. (c) Variation of the Fano mirror reflectivity with pump power. The pump power has been normalized, with 0 dB corresponding to the threshold. Here $r_B$=0 and the nanocavity is made of InP. (d) Fano laser linewidth versus pump power. The red, green, and orange curves correspond to the case where nanocavity nonlinearities are accounted for, and the nanocavity is made of InP, Si, and SiN, respectively. The black and blue curves correspond to the linear case (nanocavity nonlinearities are absent), and ${{\left| {{r}_{R}}\left( {{\omega }_{0}} \right) \right|}^{2}}$ of the blue curve is set identical to that of the red curve. Here $r_B$=0.}
\end{figure}

To simplify the situation, we assume that the nanocavity resonance is tuned so that the nonlinear resonance shift is always compensated at steady-state, i.e., $\operatorname{Re}\left\{ {{\delta }_{NL}} \right\}=0$. In practice, this can be implemented by placing electrodes close to the nanocavity for temperature or electric-field tuning. Here, considering the lack of analytical solutions and thus more time-consuming computations when including optical nonlinearities, we focus on the close-to-optimum point, $\omega_r=\omega_0$. With nanocavity nonlinearities, Fig.~\ref{fig:v3}(a), the Fano laser linewidth exhibits similar dependence on the Fano line shape as in the linear case; i.e., the narrowest linewidth is still located close to ${{R}_{\max }}$. The smallest linewidth in the linear case (black curve in Fig.~\ref{fig:v3}(b)) agrees perfectly with the approximate solution (Eq.~(\ref{eq:a4})). 

The linewidth obtained when considering optical nonlinearities is larger than the linear one and does not depend monotonically on $r_B$. This can be attributed to two factors: 1) the reduction of the Fano mirror reflectivity caused by nonlinear absorption in the nanocavity, which lowers the $Q$-factor of the laser; 2) the additional Langevin noise introduced by nanocavity nonlinearities causes additional phase fluctuations. To determine which factor is dominant, we examine ${{\left| {{r}_{R}}\left( {{\omega }_{r}} \right) \right|}^{2}}$ (see the blue curve in Fig.~\ref{fig:v3}(b)). As seen, ${{\left| {{r}_{R}}\left( {{\omega }_{r}} \right) \right|}^{2}}$ varies almost oppositely to the laser linewidth with $r_B$, indicating that the reduced Fano mirror reflectivity plays an important role. A reduced mirror reflectivity at large $r_B$ is ascribed to the fact that a higher $r_B$ can enable a larger ${{R}_{\max }}$ and thus stronger field storage in the nanocavity, resulting in higher nanocavity absorption. Indeed, the Fano mirror reflectivity decreases dramatically for high pump powers (Fig.~\ref{fig:v3}(c)). Therefore, unlike the linear case, the laser linewidth in the nonlinear case suffers from a trade-off between enhanced field localization and enhanced nanocavity absorption. This trade-off means that the linewidth does not follow the inverse-power dependence predicted by the Schawlow-Townes formula \cite{Schawlow1958} (Fig.~\ref{fig:v3}(d)). The linewidth saturates and eventually increases with increasing pump power. Simulations of other nanocavities, e.g., the H0 type \cite{Yu2013} having a smaller mode volume and a faster carrier decay rate, give qualitatively the same result. It should be noted that the nonlinear result, e.g., the red curve in Fig.~\ref{fig:v3}(d)), agrees well with the linear result when accounting for the power-dependence of ${{\left| {{r}_{R}}\left( {{\omega }_{r}} \right) \right|}^{2}}$ (blue curve in Fig.~\ref{fig:v3}(d)), further confirming that the reduction of the reflectivity due to nonlinear absorption is the dominant effect causing linewidth rebroadening. Fig.~\ref{fig:v3}(d) predicts that for an InP nanocavity, it may be difficult to reach sub-MHz linewidth. However, the problem is significantly reduced by using a material with smaller nonlinear loss, such as Si or SiN (see the green and orange curves in Fig.~\ref{fig:v3}(d)). Such structures can be realized using heterogeneous integration technology \cite{Tran2019,Crosnier2017}. We also find that nonlinear effects become orders of magnitude weaker if the nanocavity of the Fano mirror is replaced by a much larger cavity.

\section{Summary and outlook}
In summary, we have presented a general theory of the quantum-limited linewidth of a Fano laser based on a bound state in the continuum. In particular, we have developed a potential picture valid for lasers with strongly dispersive laser mirrors. Based on the theory, we show that the Fano laser allows orders-of-magnitude linewidth reduction without compromising the monostability and without significantly increasing the size of the laser. This enables microscopic lasers featuring similar linewidth-narrowing as conventional macroscopic external cavity lasers, but with different modal properties. By breaking the symmetry of the Fano mirror, we find that the Fano laser linewidth can be reduced by an additional factor of four. This improvement, however, may be compromised by optical nonlinearities that limit the minimum linewidth obtained for a given material system. 

Our theory provides new insights into the stability and coherence of microscopic lasers. The potential model accounts for global dynamics of the system that cannot be inferred from small signal analysis. The model thus facilitates the incorporation of other degrees of design freedom and physics, exemplified by the mirror symmetry breaking and optical nonlinearities considered here, which have been largely neglected in previous linewidth investigations. Therefore, the developed theory can be used to investigate other configurations, e.g., considering other dispersive laser mirrors, such as those enabled by multiple Fano resonances, Autler-Townes splitting, or electromagnetically induced transparency \cite{Limonov2017,YuOptica21,Miroshnichenko2010,Huang2022}. 

\begin{acknowledgments}
The authors acknowledge the support of the Danish National Research Foundation via NanoPhoton – Center for Nanophotonics (Grant No. DNRF147), and the European Research Council (ERC) under the European Union Horizon 2020 Research and Innovation Program (Grant No. 834410 Fano). Y. Y. acknowledges the support from Villum Fonden via the Young Investigator Program (Grant No. 42026).
\end{acknowledgments}

\appendix
\section{The effective potential of lasers with dispersive mirror}
By separating the amplitude and phase of ${{A}^{+}}\left( t \right)$ in Eq.~(\ref{eq:s1}), i.e., ${{A}^{+}}\left( t \right)=\left| {{A}^{+}}\left( t \right) \right|{{e}^{j{{\phi }^{+}}\left( t \right)}}$, we arrive at two differential equations
\begin{align}
\frac{d}{dt}\left| {{A}^{+}}\left( t \right) \right|&={{G}_{N}}\Delta N\left( t \right)\left| {{A}^{+}}\left( t \right) \right|-{{\gamma }_{in}}\left| {{A}^{+}}\left( t \right) \right| \nonumber \\
&+\operatorname{Re}\left\{ K{{A}^{-}}\left( t \right)/{{A}^{+}}\left( t \right) \right\}\left| {{A}^{+}}\left( t \right) \right|+{{F}_{\left| {{A}^{+}} \right|}}\left( t \right),\label{eq:s5}
\end{align}
\begin{align}
\frac{d}{dt}{{\phi }^{+}}\left( t \right)&=-\alpha {{G}_{N}}\Delta N\left( t \right) \nonumber \\
&+\operatorname{Im}\left\{ K{{A}^{-}}\left( t \right)/{{A}^{+}}\left( t \right) \right\}+{{F}_{{{\phi }^{+}}}}\left( t \right),\label{eq:s6}
\end{align}
where ${{F}_{\left| {{A}^{+}} \right|}}\left( t \right)$, ${{F}_{{{\phi }^{+}}}}\left( t \right)$ are the Langevin noise terms of the field amplitude and phase. The steady-state, in the absence of noise, is found by solving $d\left| {{A}^{+}}\left( t \right) \right|/dt=0$ and $d{{\phi }^{+}}\left( t \right)/dt={{\omega }_{r}}-\omega $, leading to $\alpha {{G}_{N}}\Delta N=\alpha {{\gamma }_{in}}-\alpha \operatorname{Re}\left\{ K{{r}_{R}}\left( \omega  \right) \right\}$ and
\begin{align}
{{\omega }_{r}}=\omega -\alpha {{\gamma }_{in}}+\alpha \operatorname{Re}\left\{ K{{r}_{R}}\left( \omega  \right) \right\}+\operatorname{Im}\left\{ K{{r}_{R}}\left( \omega  \right) \right\}. \label{eq:s7}
\end{align}
Here, ${{r}_{R}}\left( {{\omega }_{r}} \right)={{A}^{-}}/{{A}^{+}}$, and $\omega $ is the oscillation frequency of the entire laser system. Based on the small perturbation approach, by using ${{r}_{R}}\left( \omega  \right)=\left| {{r}_{R}}\left( \omega  \right) \right|{{e}^{j{{\phi }_{R}}\left( \omega  \right)}}$, from Eq.~(\ref{eq:s7}) one finds the following relation between changes in ${\omega }_{r}$ and ${\omega }$
\begin{align}
&\Delta {{\omega }_{r}}= \nonumber \\
&\Delta \omega +{\gamma }_{in}{{\left. \left( \alpha \frac{1}{\left| {{r}_{R}}\left( \omega  \right) \right|}\frac{\partial }{\partial \omega }\left| {{r}_{R}}\left( \omega  \right) \right|+\frac{\partial }{\partial \omega }{{\phi }_{R}}\left( \omega  \right) \right) \right|}_{{{\omega }_{r}}}}\Delta \omega. \label{eq:s8}
\end{align}
If the laser has a dispersionless mirror, i.e, $\partial \left| {{r}_{R}}\left( \omega  \right) \right|/\partial \omega ,\ \partial {{\phi }_{R}}\left( \omega  \right)/\partial \omega =0$, this reduces to the case of the equivalent FP laser, with mirrors that have the same reflectivity as the Fano laser evaluated at its operation point but with no frequency dependence, and a cavity length given by that of the Fano cavity (see Fig.~\ref{fig:v1}(a)). In this case, one has $\Delta {{\omega }_{r}}=\Delta \omega $. The factor by which the linewidth of the composite cavity laser is reduced compared to the FP laser counterpart is given by the ratio ${{\left( \Delta \omega /\Delta {{\omega }_{r}} \right)}^{2}}$ \cite{Li1989}, where $\Delta \omega $ is the change of the oscillation frequency of the composite laser system upon a change $\Delta {{\omega }_{r}}$ of the oscillation frequency of the corresponding FP laser. Using Eq.~(\ref{eq:s8}), one gets 
\begin{align}
\frac{\Delta {{v}_{FL}}}{\Delta {{v}_{FP}}}&={{\left( \frac{\Delta \omega }{\Delta {{\omega }_{r}}} \right)}^{2}}=\frac{1}{{\eta }^{2}} \nonumber \\
&=1/{{\left( 1+\frac{{{\gamma }_{in}}}{2\pi }\left( {{L}_{1}}\left( \delta  \right)+{{L}_{2}}\left( \delta  \right) \right) \right)}^{2}},\label{eq:s9}
\end{align}
where ${{L}_{1}}\left( \delta  \right)=2\pi \alpha {{\left| {{r}_{R}}\left( \omega  \right) \right|}^{-1}}{{\left. \frac{\partial }{\partial \omega }\left| {{r}_{R}}\left( \omega  \right) \right| \right|}_{\omega ={{\omega }_{r}}}}$ is the normalized amplitude derivative and ${{L}_{2}}\left( \delta  \right)=2\pi {{\left. \frac{\partial }{\partial \omega }{{\phi }_{R}}\left( \omega  \right) \right|}_{\omega ={{\omega }_{r}}}}$ is the normalized phase derivative. For the Fano laser with the Fano mirror reflectivity of Eq.~(\ref{eq:a2}), we have 
\begin{widetext}
\begin{align}
  & \frac{\partial }{\partial \omega }{{\phi }_{R}}\left( \omega  \right) \nonumber \\ 
 & =\left( -\frac{{{\delta }_{r}}\left( \omega  \right)}{{{\delta }_{r}}{{\left( \omega  \right)}^{2}}+{{\left( {{\delta }_{i}}\left( \omega  \right)-{{\gamma }_{t}} \right)}^{2}}}-\frac{r_{B}^{2}\left( P{S}_{e}-r_{B}^{2}{{\delta }_{r}}\left( \omega  \right) \right)}{{{\left( P{S}_{e}-r_{B}^{2}{{\delta }_{r}}\left( \omega  \right) \right)}^{2}}+{{\left( r_{B}^{2}\left( {{\delta }_{i}}\left( \omega  \right)-{{\gamma }_{v}} \right)-{{\gamma }_{2}}+{{\gamma }_{1}} \right)}^{2}}} \right)\frac{\partial }{\partial \omega }{{\delta }_{i}}\left( \omega  \right)\nonumber \\ 
 & +\left( \frac{{{\delta }_{i}}\left( \omega  \right)-{{\gamma }_{t}}}{{{\delta }_{r}}{{\left( \omega  \right)}^{2}}+{{\left( {{\delta }_{i}}\left( \omega  \right)-{{\gamma }_{t}} \right)}^{2}}}-\frac{r_{B}^{2}\left( r_{B}^{2}\left( {{\delta }_{i}}\left( \omega  \right)-{{\gamma }_{v}} \right)-{{\gamma }_{2}}+{{\gamma }_{1}} \right)}{{{\left( P{S}_{e}-r_{B}^{2}{{\delta }_{r}}\left( \omega  \right) \right)}^{2}}+{{\left( r_{B}^{2}\left( {{\delta }_{i}}\left( \omega  \right)-{{\gamma }_{v}} \right)-{{\gamma }_{2}}+{{\gamma }_{1}} \right)}^{2}}} \right)\frac{\partial }{\partial \omega }{{\delta }_{r}}\left( \omega  \right),\label{eq:s112} 
\end{align}
and
\begin{align}
  & {{\left| {{r}_{R}}\left( \omega  \right) \right|}^{-1}}\frac{\partial }{\partial \omega }\left| {{r}_{R}}\left( \omega  \right) \right|\nonumber \\ 
 & =\left( -\frac{{{\delta }_{r}}\left( \omega  \right)}{{{\delta }_{r}}{{\left( \omega  \right)}^{2}}+{{\left( {{\delta }_{i}}\left( \omega  \right)-{{\gamma }_{t}} \right)}^{2}}}-\frac{r_{B}^{2}\left( P{S}_{e}-r_{B}^{2}{{\delta }_{r}}\left( \omega  \right) \right)}{{{\left( P{S}_{e}-r_{B}^{2}{{\delta }_{r}}\left( \omega  \right) \right)}^{2}}+{{\left( r_{B}^{2}\left( {{\delta }_{i}}\left( \omega  \right)-{{\gamma }_{v}} \right)-{{\gamma }_{2}}+{{\gamma }_{1}} \right)}^{2}}} \right)\frac{\partial }{\partial \omega }{{\delta }_{r}}\left( \omega  \right)\nonumber \\ 
 & -\left( \frac{{{\delta }_{i}}\left( \omega  \right)-{{\gamma }_{t}}}{{{\delta }_{r}}{{\left( \omega  \right)}^{2}}+{{\left( {{\delta }_{i}}\left( \omega  \right)-{{\gamma }_{t}} \right)}^{2}}}-\frac{r_{B}^{2}\left( r_{B}^{2}\left( {{\delta }_{i}}\left( \omega  \right)-{{\gamma }_{v}} \right)-{{\gamma }_{2}}+{{\gamma }_{1}} \right)}{{{\left( P{S}_{e}-r_{B}^{2}{{\delta }_{r}}\left( \omega  \right) \right)}^{2}}+{{\left( r_{B}^{2}\left( {{\delta }_{i}}\left( \omega  \right)-{{\gamma }_{v}} \right)-{{\gamma }_{2}}+{{\gamma }_{1}} \right)}^{2}}} \right)\frac{\partial }{\partial \omega }{{\delta }_{i}}\left( \omega  \right),\label{eq:s122} 
\end{align}
\end{widetext}
where ${{\delta }_{r}}\left( \omega  \right)$ (${{\delta }_{i}}\left( \omega  \right)$) is the real (imaginary) part of $\delta +{{\delta }_{NL}}\left( \omega  \right)$, and its specific form depends on the type of optical nonlinearities in the nanocavity, ${\delta }_{NL}\left( \omega  \right)$.

In the linear case where ${{\delta }_{NL}}=0$ (${{\delta }_{r}}\left( \omega  \right)={{\omega }_{0}}-\omega ={{\omega }_{0}}-{{\omega }_{r}}$, ${{\delta }_{i}}\left( \omega  \right)=0$), the above equations reduce to
\begin{equation}
\frac{\partial }{\partial \omega }{{\phi }_{R}}\left( \omega  \right)=\frac{{{\gamma }_{t}}}{{{\delta }^{2}}+\gamma _{t}^{2}}-\frac{r_{B}^{2}\left( r_{B}^{2}{{\gamma }_{v}}+ {{\gamma }_{2}}-{{\gamma }_{1}}  \right)}{{{\left( P{S}_{e}-r_{B}^{2}\delta  \right)}^{2}}+{{\left( r_{B}^{2}{{\gamma }_{v}}+ {{\gamma }_{2}}-{{\gamma }_{1}}  \right)}^{2}}},\nonumber
\end{equation}
and
\begin{align}
&{{\left| {{r}_{R}}\left( \omega  \right) \right|}^{-1}}\frac{\partial }{\partial \omega }\left| {{r}_{R}}\left( \omega  \right) \right| \nonumber \\
&=\frac{\delta }{{{\delta }^{2}}+\gamma _{t}^{2}}+\frac{r_{B}^{2}\left( P{S}_{e}-r_{B}^{2}\delta  \right)}{{{\left( P{S}_{e}-r_{B}^{2}\delta  \right)}^{2}}+{{\left( r_{B}^{2}{{\gamma }_{v}}+ {{\gamma }_{2}}-{{\gamma }_{1}} \right)}^{2}}},\label{eq:s11}
\end{align}
in which ${S}_{e}={{t}_{B}}\sqrt{4{{\gamma }_{1}}{{\gamma }_{2}}-t_{B}^{2}{{\left( {{\gamma }_{1}}+{{\gamma }_{2}} \right)}^{2}}}$. The laser stability can be investigated through a conventional small-signal analysis, or investigated in the time-domain, where we solve Eqs.~(\ref{eq:s1})-(\ref{eq:s4}) numerically by treating the Langevin noise terms as random sources with normal distribution \cite{Ahmed2001}.
From Eqs.~(\ref{eq:s5}) and (\ref{eq:s6}), and neglecting amplitude fluctuations, one gets
\begin{align}
& \frac{d}{dt}{{\phi }^{+}}\left( t \right) \nonumber \\ 
& =-\alpha \left( {{\gamma }_{in}}-\operatorname{Re}\left\{ K{{r}_{R}}\left( \omega  \right) \right\} \right)+\operatorname{Im}\left\{ K{{r}_{R}}\left( \omega  \right) \right\}+{{F}_{{{\phi }^{+}}}}\left( t \right) \nonumber \\ 
& =-\alpha {{\gamma }_{in}}+{{\gamma }_{in}}\left| \frac{{{r}_{R}}\left( \omega  \right)}{{{r}_{R}}\left( {{\omega }_{r}} \right)} \right|(\alpha \cos \left( {{\phi }_{R}}\left( \omega  \right)-{{\phi }_{R}}\left( {{\omega }_{r}} \right) \right) \nonumber \\ 
& \ \ \ \ \ \ \ \ \ \ \ \ +\sin \left( {{\phi }_{R}}\left( \omega  \right)-{{\phi }_{R}}\left( {{\omega }_{r}} \right) \right))+{{F}_{{{\phi }^{+}}}}\left( t \right). \label{eq:s12} 
\end{align}

\onecolumngrid
\begin{center}
\setlength{\abovecaptionskip}{3pt} % Chosen fairly arbitrarily
\begin{figure}[h]
\includegraphics[scale=0.4]{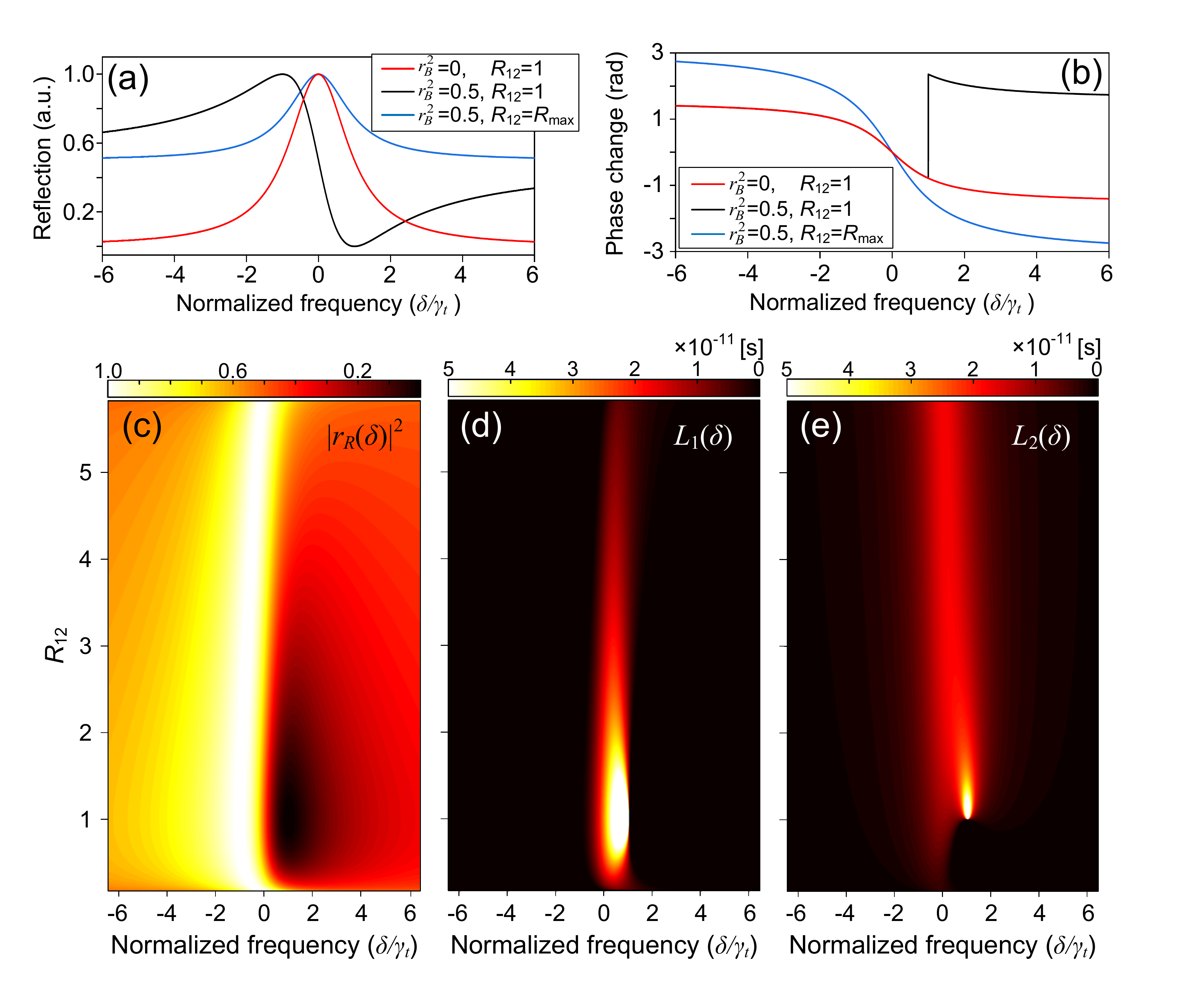}
%\begin{figure}
%\centering
%\includegraphics[width=1\linewidth]{FigS1.pdf}
\caption{\label{fig:s1} (a) The Fano mirror reflectivity (${{\left| {{r}_{R}}\left( \delta  \right) \right|}^{2}}$) for different reflectivities of the PTE ($r_{B}^{2}$) and different decay ratios (${{R}_{12}}={{\gamma }_{1}}/{{\gamma }_{2}}$).  Here, $\delta ={{\omega }_{0}}-\omega $, with ${{\omega }_{0}}$ being the nanocavity resonant frequency. (b) The phase change of the Fano mirror reflection ($\arg \left\{ {{r}_{R}}\left( \delta  \right) \right\}$) for different reflectivities of the PTE and the decay ratios (the phase change was normalized so that $\arg \left\{ {{r}_{R}}\left( 0 \right) \right\}=0$). (c) The mirror reflectivity ${{\left| {{r}_{R}}\left( \delta  \right) \right|}^{2}}$, (d) normalized amplitude derivative ${{L}_{1}}\left( \delta  \right)$, and (e) normalized phase derivative ${{L}_{2}}\left( \delta  \right)$ of the Fano mirror, as a function of $\delta /{{\gamma }_{t}}$ and ${{R}_{12}}$. Here $r_{B}^{2}$=0.5 and $\alpha $=3. In (d) and (e), the ranges of color scales are restricted to the interval from 0 to 5$\times $10$^{-11}$ s for a clearer illustration.}
\end{figure}
\end{center}
\twocolumngrid

Next, we extend the approach taken in Ref.~\cite{MORK1992E} for external cavity lasers to the general scenario. We introduce ${{\phi }^{+}}\left( t-{{\tau }_{D}} \right)$ which is the phase delayed by a time, $\tau_{D}$. Such a delay is caused by field dwelling in the external cavity in conventional external cavity lasers or field storage in the nanocavity in the Fano laser. For example, ${{\tau }_{D}}=1/{{\gamma }_{D}}=\partial {{\phi }_{R}}\left( \omega  \right)/\partial \omega =1/{{\gamma }_{t}}$ is the time delay at ${{\omega }_{r}}={{\omega }_{0}}$ when ${{\gamma }_{v}}=0$ and ${{\gamma }_{2}}={{\gamma }_{1}}$. By defining $\Delta ={{\phi }^{+}}\left( t \right)-{{\phi }^{+}}\left( t-{{\tau }_{D}} \right)$, using $\Delta \approx \left( {{\omega }_{r}}-\omega  \right){{\tau }_{D}}$ and $\left\langle \partial {{\phi }^{+}}\left( t-{{\tau }_{D}} \right)/\partial t \right\rangle \approx {{\gamma }_{D}}\Delta $, we derive from Eq.~(\ref{eq:s12}) the following equation for $\Delta $ 
\begin{align}
\frac{d}{dt}\Delta &=\frac{d}{dt}{{\phi }^{+}}\left( t \right)-\frac{d}{dt}{{\phi }^{+}}\left( t-{{\tau }_{D}} \right) \nonumber \\ 
&=\frac{d}{dt}{{\phi }^{+}}\left( t \right)-\left( \left\langle \frac{d}{dt}{{\phi }^{+}}\left( t-{{\tau }_{D}} \right) \right\rangle +{{F}_{{{\phi }^{+}}}}\left( t-{{\tau }_{D}} \right) \right) \nonumber \\
&=-\frac{dV}{d\Delta }+{{F}_{\Delta }}\left( t \right),\label{eq:s13}
\end{align}
where
\begin{align}
\frac{dV}{d\Delta }&=\Delta {\gamma_D}+\alpha {\gamma_{in}} \nonumber \\
&-{\gamma_{in}}\left| {r_R}\left( \omega  \right)/{r_R}\left( {\omega_r} \right) \right| ( \alpha \cos \left( {\phi_R}\left( \omega  \right)-{\phi_R}\left( {\omega_r} \right) \right) \nonumber \\
&+\sin \left( {\phi_R}\left( \omega  \right)-{\phi_R}\left( {\omega_r} \right) \right),\label{eq:s14}
\end{align}
and ${F_{\Delta }}\left( t \right)={F_{\phi^+}}\left( t \right)-{F_{\phi }^{+}}\left( t-{{\tau }_D} \right)$ is a Langevin term with a correlation strength $\left\langle {F_{\Delta }}\left( t \right){F_{\Delta }}\left( t' \right) \right\rangle =4\pi \Delta {v_{FP}}\left( {{\omega }_{r}} \right)\delta \left( t-t' \right)$. Eq.~(\ref{eq:s13}) is analogous to the equation of motion for a particle with coordinate $\Delta $ moving with strong friction in a potential $V$ and exposed to a fluctuating force ${F_{\Delta }}\left( t \right)$ \cite{Mark1992}. The coordinate $\Delta $ plays the role of the instantaneous laser frequency as well as the phase change over the time interval ${{\tau }_D}$. Taking the Fano mirror reflectivity (Eq.~(\ref{eq:a2})) and considering the simple case of ${{\gamma }_1}={{\gamma }_2}={{\gamma }_c}/2\approx {{\gamma }_t}/2$, we obtain by integrating Eq.~(\ref{eq:s14}) 
\begin{widetext}
\begin{align}
V&=\frac{1}{2}{{\gamma }_{D}}{{\Delta }^{2}}+\alpha {{\gamma }_{in}}\Delta \nonumber \\
& -\frac{{{\gamma }_{in}}}{2{{\gamma }_{D}}}\frac{2{{r}_{B}}\left( {{\gamma }_{D}}\Delta +{{\delta }} \right)\left( {{\gamma }_{t}}-\alpha {{\delta }} \right)+2{{\gamma }_{t}}\left( {{t}_{B}}\left( \alpha {{\gamma }_{t}}+{{\delta }} \right)-{{r}_{B}}\left( {{\gamma }_{t}}-\alpha {{\delta }} \right) \right)\tan^{-1}\left( \frac{{{\gamma }_{D}}\Delta +{{\delta }}}{{{\gamma }_{t}}} \right)}{{{t}_{B}}{{\gamma }_{t}}-{{r}_{B}}{{\delta }}}\nonumber \\
& +\frac{{{\gamma }_{in}}}{2{{\gamma }_{D}}}\frac{{{\gamma }_{t}}\left( {{r}_{B}}\left( \alpha {{\gamma }_{t}}+{{\delta }} \right)+{{t}_{B}}\left( {{\gamma }_{t}}-\alpha {{\delta }} \right) \right)\ln \left\{ \gamma _{t}^{2}+{{\left( {{\gamma }_{D}}\Delta +{{\delta }} \right)}^{2}} \right\}}{{{t}_{B}}{{\gamma }_{t}}-{{r}_{B}}{{\delta }}},\label{eq:s15} 
\end{align}
\end{widetext}
where ${{\gamma }_{D}}=1/\left( \partial {{\phi }_{R}}\left( \omega  \right)/\partial \omega  \right)$. For $\alpha =0$, ${{r}_{B}}=0$ (${{t}_{B}}=1$), and ${{\omega }_{r}}={{\omega }_{0}}$, we have ${{\gamma }_{D}}={{\gamma }_{t}}$, so Eq.~(\ref{eq:s15}) reduces to
\begin{equation}
\ \ \ \ \ \ V=\frac{1}{2}{{\gamma }_{D}}{{\Delta }^{2}}+{{\gamma }_{in}}\frac{1}{2}\ln \left\{ 1+{{\Delta }^{2}} \right\}. \label{eq:s16}
\end{equation}
The potential Eq.~(\ref{eq:s16}) for the Fano laser is different from the potential for an external cavity laser \cite{Mark1992}, which (for $\alpha =0$) assumes the form 
\begin{equation}
V=\frac{1}{2}{{\gamma }_{D}}{{\Delta }^{2}}-\kappa {{\gamma }_{in}}\cos \left( {{\theta }_{0}}+\Delta  \right). \label{eq:s17}
\end{equation}
From Eq.~(\ref{eq:s14}), by considering $\partial \omega /\partial \Delta =-{{\gamma }_{D}}$ when $\omega \to {{\omega }_{r}}$, we find the curvature of the potential $V$ in the minimum point 
\begin{align}
&{{\left. \frac{1}{{{\gamma }_{D}}}\frac{{{d}^{2}}V}{d{{\Delta }^{2}}} \right|}_{\omega ={{\omega }_{r}}}} \nonumber\\
&\approx 1+{{\gamma }_{in}}{{\left. \left( \alpha {{\left| {{r}_{R}}\left( \omega  \right) \right|}^{-1}}\frac{\partial }{\partial \omega }\left| {{r}_{R}}\left( \omega  \right) \right|+\frac{\partial }{\partial \omega }{{\phi }_{R}}\left( \omega  \right) \right) \right|}_{\omega={{\omega }_{r}}}} \nonumber\\
&=\eta ={{\left( \frac{\Delta {{v}_{FP}}}{\Delta {{v}_{FL}}} \right)}^{\frac{1}{2}}}.\label{eq:s18}
\end{align}
Eq.~(\ref{eq:s18}) shows that the curvature of $V$ at $\omega ={{\omega }_{r}}$, normalized by ${{\gamma }_{D}}$, is identical to the linewidth reduction factor for the Fano laser compared to the FP laser counterpart.

\section{Coupled-cavity system versus the Fano system with partially transmitting element}
For simplicity, we consider a Fano laser system with PTE, where ${{r}_{B}}$=1 (${{t}_{B}}$=0) and $R_{12}=R_{max}$. We neglect the linewidth enhancement factor and optical nonlinearities in the nanocavity. By replacing ${{A}_{FP}}\left( t \right)=j{{A}^{+}}\left( t \right)/\sqrt{{{\gamma }_{in}}}$, and assuming that the laser oscillates at a frequency where the Fano mirror reflectivity is close to unity ($\left| {{r}_{R}}\left( {{\omega }_{r}} \right) \right|\to 1$ with ${{\gamma }_{v}}\to 0$) so that the inverse of the photon lifetime of the Fano cavity can be simplified as ${{\gamma }_{p}}=2{{\gamma }_{i}}+2{{\gamma }_{in}}\ln \left\{ 1/\left| {{r}_{R}}\left( {{\omega }_{r}} \right) \right| \right\}\approx 2{{\gamma }_{i}}+2{{\gamma }_{in}}\left( 1-\left| {{r}_{R}}\left( {{\omega }_{r}} \right) \right| \right)$, Eqs.~(\ref{eq:s1}) and (\ref{eq:s4}) can be reduced, in a matrix form, to
\begin{equation}
\frac{d}{dt}\left( \begin{matrix}
   {{A}_{c}}\left( t \right)  \\
   {{A}_{FP}}\left( t \right)  \\
\end{matrix} \right)=\mathbf{M}\left( \begin{matrix}
   {{A}_{c}}\left( t \right)  \\
   {{A}_{FP}}\left( t \right)  \\
\end{matrix} \right), \label{eq:s118}
\end{equation}
where
\begin{equation}
\mathbf{M}=\left( \begin{matrix}
   -j{{\delta }_{0}}-{{\gamma }_{t}} & {{\kappa }_{c}}  \\
   {{\kappa }_{c}} & g-{{\gamma }_{in}}  \\
\end{matrix} \right). \nonumber
\end{equation}
Here $g={{G}_{N}}\left( N-{{N}_{0}} \right)-{{\gamma }_{i}}$, and ${{\kappa }_{c}}$ is, in general, complex but approaches $\sqrt{2{{\gamma }_{in}}{{\gamma }_{c}}}$ for ${{\gamma }_{v}}\ll{{\gamma }_{c}}$. By using $\left| {{r}_{R}}\left( {{\omega }_{r}} \right) \right|\to 1$ and ${{G}_{N}}\left( N-{{N}_{0}} \right)\to {{\gamma }_{i}}$, Eq.~(\ref{eq:s118}) leads to two eigenfrequencies
\begin{align}
  & {{\omega }_{e,1}}={{\omega }_{0}}-\frac{1}{2}({{\delta }_{0}}-j\left( 2{{\gamma }_{in}}+{{\gamma }_{t}} \right) \nonumber\\ 
 & \ \ \ \ \ \ \ \ \ +j\sqrt{8{{\gamma }_{in}}{{\gamma }_{c}}-{{\left( {{\delta }_{0}}+j\left( 2{{\gamma }_{in}}-{{\gamma }_{t}} \right) \right)}^{2}}}), \nonumber\\ 
 & {{\omega }_{e,2}}={{\omega }_{0}}-\frac{1}{2}({{\delta }_{0}}-j\left( 2{{\gamma }_{in}}+{{\gamma }_{t}} \right) \nonumber\\ 
 & \ \ \ \ \ \ \ \ \ -j\sqrt{8{{\gamma }_{in}}{{\gamma }_{c}}-{{\left( {{\delta }_{0}}+j\left( 2{{\gamma }_{in}}-{{\gamma }_{t}} \right) \right)}^{2}}}). \nonumber 
\end{align}
We can get the $Q$-factors of the eigenmodes of the Fano laser system as ${{Q}_{e,1/2}}=-\operatorname{Re}\left\{ {{\omega }_{e,1/2}} \right\}/\left( 2\operatorname{Im}\left\{ {{\omega }_{e,1/2}} \right\} \right)$, and for ${{\delta }_{0}}=0$, we get ${{Q}_{e,1}}=\infty $, ${{Q}_{e,2}}={{\omega }_{0}}/\left( 2\left( 2{{\gamma }_{in}}+{{\gamma }_{c}} \right) \right)$. The corresponding eigenvectors are
\begin{equation}
{\mathbf{v}_{e,1}}=\left( \begin{matrix}
   \sqrt{2{{\gamma }_{in}}/{{\gamma }_{t}}}  \\
   1  \\
\end{matrix} \right),
{\mathbf{v}_{e,2}}=\left( \begin{matrix}
   -\sqrt{{{\gamma }_{t}}/\left( 2{{\gamma }_{in}} \right)}  \\
   1  \\
\end{matrix} \right).
\nonumber 
\end{equation}
As seen, the eigenmode 1, with ${{Q}_{e,1}}=\infty $, corresponds to the Fano mode where the field is concentrated in the nanocavity (since ${{\gamma }_{in}}\gg{{\gamma }_{t}}$). This agrees with our previous analysis \cite{Yu2021}. The eigenmode 2 corresponds to the case where the field is concentrated in the WG part. Interestingly, the eigenmode 2 has a low $Q$-factor even though the reflectivity of the PTE is unity, i.e., the WG is closed at the right end. Such a low $Q$-factor can be ascribed to the fact that when the field is concentrated in the WG part while the nanocavity is almost empty, the phase change of the WG field induced by the reflection off the right Fano mirror has a phase difference of $\pi$ compared to the case where the nanocavity field is well established. This means that eigenmode 2 dose not meet the resonant condition of the FP cavity defined by the left end of the WG and the right PTE. So the field dissipates quickly. Therefore, the Fano mode can be still classified as a BIC even when the PTE completely blockes the WG. 

From Eq.~(\ref{eq:s118}), since ${{\kappa }_{c}}$ is almost purely real for the Fano laser, it contrasts with ordinary coupled-cavity systems where ${{\kappa }_{c}}$ is imaginary \cite{Haus}. This distinguishes the Fano system from the case of Autler-Townes splitting \cite{ATAT1955}, which corresponds to a mode doublet with similar $Q$-factors. Our system rather bears resemblance to the Parity-Time system \cite{Hodaei2014,Feng2014} working in the Parity-Time broken regime where the two eigenmodes split in loss. However, a fundamental difference is that the lasing mode of the Fano laser is the eigenmode 1, with the field concentrated in the passive (low loss) nanocavity region, while it is distributed evenly between the passive and active regions in the Parity-Time symmetric regime, or concentrated in the active (high loss) region in the Parity-Time broken regime. This is because a real $\kappa_{c} $ in the Fano laser enables the loss of one cavity to be compensated by the feedback from the other one, i.e., the lasing is promoted by decreasing the “mirror” loss of the mode through field destructive interference between ${{A}_{c}}$ (discrete mode) and ${{A}_{FP}}$ (quasi-continuum mode). For the Parity-Time laser (in the Parity-Time broken regime), instead, the loss is compensated by an enhanced modal gain as in that regime, $\left| {{\kappa }_{c}} \right|$ is usually small. A smaller $\left| {{\kappa }_{c}} \right|$ will localize a larger portion of the field in the active region, i.e., the lasing is promoted by increasing the lateral optical confinement factor. It should be noted that the BIC of our Fano system can transit to a conventional coupled-cavity system when ${{t}_{B}}$=0 and ${{\gamma }_{v}}\gg{{\gamma }_{c}}$. In this case, ${{\kappa }_{c}}\to j{{\kappa }_{c}}$, i.e., ${{\kappa }_{c}}$ becomes imaginary, which means in practice that the nanocavity and WG have a large spatial separation.

\section{Laser linewidth based on the Langevin approach}
Based on Eqs.~(\ref{eq:s1})-(\ref{eq:s4}) and Eqs.~(\ref{eq:a10}) and (\ref{eq:a11}), neglecting the Langevin noise forces, , we obtain the steady-state solutions of the Fano laser system (for ${{\omega }_{r}}={{\omega }_{0}}$) by solving the following set of algebraic equations
\begin{subequations} 
\begin{align}
&{{N}_{c}}={{G}_{T}}{{\left| {{A}_{c}} \right|}^{4}}/{{\gamma }_{nc}}, \nonumber\\
&{{r}_{R}}\left( {{\omega }_{0}} \right)={{r}_{B}}+\frac{2{{\gamma }_{1}}{{e}^{2j{{\theta }_{1}}}}}{j{{\delta }_{0}}+{{\gamma }_{t}}+{{K}_{T}}{{\left| {{A}_{c}} \right|}^{2}}+U{{\left| {{A}_{c}} \right|}^{4}}}, \nonumber \\
&N={{N}_{0}}+\left( {{\gamma }_{i}}+{{\gamma }_{in}}\ln \left\{ 1/\left| {{r}_{R}}\left( {{\omega }_{0}} \right) \right| \right\} \right)/{{G}_{N}},\nonumber\\ 
&\left| {{A}^{+}} \right|\nonumber\\
&=\sqrt{\frac{\hbar {{\omega }_{0}}{{V}_{a}}\left( R-{{\gamma }_{N}}N \right)}{\left( 1-{{\left| {{r}_{R}}\left( {{\omega }_{0}} \right) \right|}^{2}} \right)\left( {{\gamma }_{i}}/\left( {{\gamma }_{in}}\ln \left\{ 1/\left| {{r}_{R}}\left( {{\omega }_{0}} \right) \right| \right\} \right)+1 \right)}}, \nonumber\\
&{{A}^{-}}={{r}_{R}}\left( {{\omega }_{0}} \right)\left| {{A}^{+}} \right|,\    {{A}_{c}}=\left( {{A}^{-}}-{{r}_{B}}\left| {{A}^{+}} \right| \right)/\sqrt{2{{\gamma }_{1}}}, \nonumber\\
&{{\phi }^{+}}=0, \ {{\phi }^{-}}=-j\ln \left( {{A}_{-}}/\left| {{A}_{-}} \right| \right), \nonumber\\
&{{\phi }_{c}}=-j\ln \left( {{A}_{c}}/\left| {{A}_{c}} \right| \right). \nonumber  
\end{align}
\end{subequations}
Here, we define $U={{G}_{T}}{{K}_{A}}/{{\gamma }_{nc}}$, and ${{\left| {{A}_{c}} \right|}^{2}}$ is the real and positive solution ($X$) of the following equation
\begin{align}
&{{U}^{2}}{{X}^{5}}+2U{{K}_{T}}{{X}^{4}}+\left( K_{T}^{2}+2U{{\gamma }_{t}} \right){{X}^{3}}+2{{K}_{T}}{{\gamma }_{t}}{{X}^{2}}\nonumber\\
&+\left( \delta _{0}^{2}+\gamma _{t}^{2} \right)X-2{{\gamma }_{1}}{{\left| {{A}^{+}} \right|}^{2}}=0.
\end{align}
We assume that the Fano resonance can be tuned so that the nonlinear resonance shift of the nanocavity is always compensated at steady-state, i.e., $\operatorname{Re}\left\{ {{\delta }_{NL}} \right\}=0$. Next, we separate the amplitude and phase of the fields, i.e., ${{A}^{\pm }}\left( t \right)=\left| {{A}^{\pm }}\left( t \right) \right|{{e}^{j{{\phi }^{\pm }}\left( t \right)}}$, ${{A}_{c}}\left( t \right)=\left| {{A}_{c}}\left( t \right) \right|{{e}^{j{{\phi }_{c}}\left( t \right)}}$, expand the perturbation of the dynamical variables to first-order around their steady states, i.e., $\mathbf{H}\left( t \right)=\mathbf{H}+\Delta \mathbf{H}\left( t \right)$ with $\mathbf{H}={{\left[ \left| {{A}^{+}} \right|,\ \left| {{A}^{-}} \right|,\ \left| {{A}_{c}} \right|,\ {{\phi }^{+}},\ {{\phi }^{-}},\ \ {{\phi }_{c}},\ N,\ {{N}_{c}} \right]}^{T}}$ being the steady-state values. After that, by Fourier transforming the perturbations to the frequency domain, we arrive at the relation
\begin{equation}
    \mathbf{O}\Delta \mathbf{H}\left( \omega  \right)=\mathbf{F}\left( \omega  \right). \label{eq:s27}
\end{equation}
Here, $\mathbf{O}$ is a coefficient matrix depending on the laser parameters, the nonlinear coefficients, and steady-state values \cite{Coldren}. $\mathbf{F}\left( \omega  \right)$ is the Langevin noise terms [$F_{\left| A^+ \right|}\left( \omega  \right)$, 0, ${F_{\left| {A_c} \right|}}\left( \omega  \right)$, $F_{\phi^+}\left( \omega  \right)$, 0, $F_{\phi_c}\left( \omega  \right)$,  $F_N\left( \omega  \right)$, 0]$^T$ whose correlation strengths can be evaluated by inspecting the average particle exchange rates into and out of various reservoirs \cite{Coldren}. For simplicity, if neglecting the shot noise associated with nanocavity resonance shift, after some algebra, we get
\begin{subequations}
\begin{align}
&\left\langle {{F}_{\left| {{A}^{+}} \right|}}{{F}_{\left| {{A}^{+}} \right|}} \right\rangle ={{G}_{N}}\left( N-{{N}_{0}} \right){{n}_{sp}}/{{\varsigma }_{s}}\left( {{\omega }_{0}} \right),\nonumber\\  
&\left\langle {{F}_{{{\phi }^{+}}}}{{F}_{{{\phi }^{+}}}} \right\rangle =\left\langle {{F}_{\left| {{A}^{+}} \right|}}{{F}_{\left| {{A}^{+}} \right|}} \right\rangle /{{\left| {{A}^{+}} \right|}^{2}},\nonumber\\
&\left\langle {{F}_{N}}{{F}_{N}} \right\rangle =4{{G}_{N}}\left( N-{{N}_{0}} \right){{\varsigma }_{s}}\left( {{\omega }_{0}} \right){{\left| {{A}^{+}} \right|}^{2}}\left( {{n}_{sp}}-1 \right)/V_{c}^{2}\nonumber\\
& \ \ \ \ \ \ \ \ \ \ \ +2R/{{V}_{c}},\nonumber\\
&\left\langle {{F}_{\left| {{A}^{+}} \right|}}{{F}_{N}} \right\rangle \nonumber\\
&=\frac{{{G}_{N}}\left( N-{{N}_{0}} \right)\left( {{\varsigma }_{s}}\left( {{\omega }_{0}} \right)\left( 1-2{{n}_{sp}} \right){{\left| {{A}^{+}} \right|}^{2}}-{{n}_{sp}}\right)} {{{\varsigma }_{s}}\left( {{\omega }_{0}} \right){{V}_{c}}\left| {{A}^{+}} \right|},\nonumber\\
&\left\langle {{F}_{\left| {{A}_{c}} \right|}}{{F}_{\left| {{A}_{c}} \right|}} \right\rangle =\hbar {{\omega }_{0}}\left( {{K}_{A}}{{N}_{c}}+{{K}_{T}}{{\left| {{A}_{c}} \right|}^{2}}+{{\gamma }_{v}} \right)/2,\nonumber\\  
&\left\langle {{F}_{{{\phi }_{c}}}}{{F}_{{{\phi }_{c}}}} \right\rangle =\left\langle {{F}_{\left| {{A}_{c}} \right|}}{{F}_{\left| {{A}_{c}} \right|}} \right\rangle /{{\left| {{A}_{c}} \right|}^{2}}.\nonumber
\end{align}
\end{subequations}
Here $\left\langle {} \right\rangle $ indicates a statistical ensemble average, which is identical to a time average in the present case of an ergodic system. We have neglected the Langevin noise terms due to the free carriers generated in the nanocavity. The phase $\Delta {{\phi }_{c}}\left( \omega  \right)$ can be obtained by solving Eq.~(\ref{eq:s27}), and the laser frequency fluctuation is ${{v}_{c}}\left( \omega  \right)=-j\omega \Delta {{\phi }_{c}}\left( \omega  \right)/\left( 2\pi  \right)$, which can be expressed analytically in terms of the Langevin noise sources
\begin{align}
    &{{v}_{c}}\left( \omega  \right)={{\zeta }_{1}}{{F}_{\left| {{A}^{+}} \right|}}\left( \omega  \right)+{{\zeta }_{2}}{{F}_{\left| {{A}_{c}} \right|}}\left( \omega  \right)+{{\zeta }_{3}}{{F}_{{{\phi }^{+}}}}\left( \omega  \right)\nonumber \\
    &+{{\zeta }_{4}}{{F}_{{{\phi }_{c}}}}\left( \omega  \right)+{{\zeta }_{5}}{{F}_{N}}\left( \omega  \right).\nonumber
\end{align}
The noise frequency spectrum can thus be obtained as
\begin{align}
& {{S}_{v}}\left( \omega  \right)=\frac{1}{2\pi }\int{\left\langle {{v}_{c}}\left( \omega  \right){{v}_{c}}{{\left( {{\omega }'} \right)}^{*}} \right\rangle d{\omega }'}\nonumber \\ 
& ={{\left| {{\zeta }_{1}} \right|}^{2}}\left\langle {{F}_{\left| {{A}^{+}} \right|}}{{F}_{\left| {{A}^{+}} \right|}} \right\rangle +{{\left| {{\zeta }_{2}} \right|}^{2}}\left\langle {{F}_{\left| {{A}_{c}} \right|}}{{F}_{\left| {{A}_{c}} \right|}} \right\rangle \nonumber \\
& +{{\left| {{\zeta }_{3}} \right|}^{2}}\left\langle {{F}_{{{\phi }^{+}}}}{{F}_{{{\phi }^{+}}}} \right\rangle +{{\left| {{\zeta }_{4}} \right|}^{2}}\left\langle {{F}_{{{\phi }_{c}}}}{{F}_{{{\phi }_{c}}}} \right\rangle +{{\left| {{\zeta }_{5}} \right|}^{2}}\left\langle {{F}_{N}}{{F}_{N}} \right\rangle \nonumber \\
& +\left( {{\zeta }_{1}}\zeta _{5}^{*}+\zeta _{1}^{*}{{\zeta }_{5}} \right)\left\langle {{F}_{\left| {{A}^{+}} \right|}}{{F}_{N}} \right\rangle, \label{eq:s28}  
\end{align}
which depends on the Langevin noise correlation strengths. The laser linewidth is finally found as
\begin{equation}
    \Delta {{v}_{FL}}=2\pi {{S}_{v}}\left( 0 \right).
\end{equation}
Noting that by neglecting the Langevin noise correlation terms in Eq.~(\ref{eq:s28}), except the dominating terms, $\left\langle {{F}_{\left| {{A}^{+}} \right|}}{{F}_{\left| {{A}^{+}} \right|}} \right\rangle $ and $\left\langle {{F}_{{{\phi }^{+}}}}{{F}_{{{\phi }^{+}}}} \right\rangle $, the expression for the Fano laser linewidth reduces to Eq.~(\ref{eq:a3}) in the main text. Here, the nonlinear coefficients ${{K}_{K}}$, ${{K}_{T}}$, ${{K}_{D}}$, ${{K}_{A}}$, ${{G}_{T}}$, obtained based on the calculated nanocavity nonlinear mode volumes combined with material parameters \cite{Yu2013,Gunter1987,Sohn2019}, are listed in TABLE I. 
\begin{table}
\begin{ruledtabular}
\caption{Nonlinear coefficients for the simulations}\label{tab:params}
\begin{tabular} {c c c c}
\textbf{Coefficients}              & \textbf{InP} &\textbf{Si} &\textbf{SiN} \\\hline 
${{K}_{K}}$ (W$^{-1}$s$^{-2}$)      & -8.8$\times$10$^{23}$ & -1.2$\times$10$^{23}$ & -1.2$\times$10$^{23}$\\   
${{K}_{T}}$ (W$^{-1}$s$^{-2}$)     & 1.85$\times$10$^{24}$ & 2.9$\times$10$^{22}$ & 6.6$\times$10$^{21}$\\
${{K}_{D}}$ (m$^{3}$/s)     & 1.95$\times$10$^{-12}$ & 4.8$\times$10$^{-13}$ & 0\\   
${{K}_{A}}$ (m$^{3}$/s)    & 2.13$\times$10$^{-13}$ & 6.9$\times$10$^{-14}$ & 0\\
${{G}_{T}}$ (W$^{-2}$m$^{-3}$s$^{-3}$)      & 3.35$\times$10$^{61}$ & 5.3$\times$10$^{59}$ & 1.2$\times$10$^{59}$\\
\end{tabular}
\end{ruledtabular}
\end{table}

\section{Output power of the Fano laser}
When the Fano laser operates around the nanocavity resonance, the output power transmitted through the Fano mirror (termed as through-port) is limited due to the very high reflectivity of the Fano mirror. 
\begin{figure}
\centering
\includegraphics[width=1.0\linewidth]{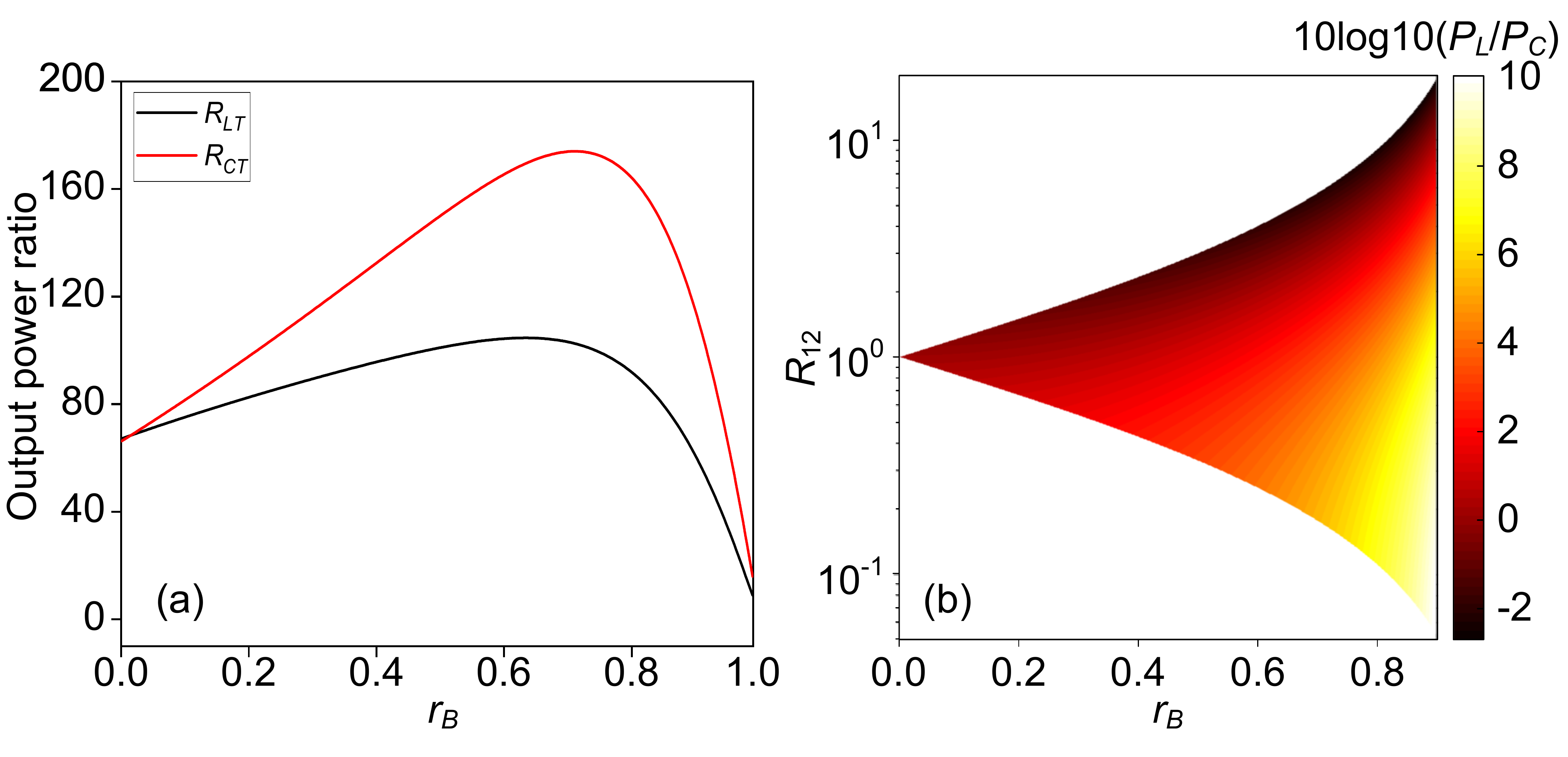}
\caption{\label{fig:s2} (a) The ratio of output powers in the left port and the through-port (${{R}_{LT}}$, black curve) and the ratio of output powers in the cross- and through-port (${{R}_{CT}}$, red curve), as a function of $r_B$ when the Fano laser operates at the frequency point of the optimum linewidth ${{\omega }_{m}}$ (as $R_{12}$=$R_{max}$). (b) The output power ratio (log scale) between the left- and cross-port as a function of $r_B$ and $R_{12}$. The Fano laser operates at the reflectivity peak of the Fano mirror.}
\end{figure}
Therefore, it is preferable to use another channel for the output, for example, e.g., either the left mirror (the left end of the WG) by reducing (increasing) the left mirror reflectivity (transitivity), $r_L$ ($t_L$), or through a cross-port by placing an additional WG adjacent to the nanocavity with a coupling efficiency ${{\gamma }_{3}}$ \cite{Mork2019}. These channels are expected to have a higher external quantum efficiency than the through-port \cite{Mork2019}. The ratio of the output power of the left mirror, $P_L$, with respect to the through-port, $P_T$, can be obtained as
\begin{equation}
    {{R}_{LT}}\left( {{\omega }_{r}} \right)=\frac{{{P}_{L}}}{{{P}_{T}}}=\frac{\left( 1-r_{L}^{2} \right)\left| {{r}_{R}}\left( {{\omega }_{r}} \right) \right|}{{{r}_{L}}\left( 1-{{\left| {{r}_{R}}\left( {{\omega }_{r}} \right) \right|}^{2}} \right)}, \nonumber
\end{equation}
and the ratio of the output power of the cross-port, $P_C$, with respect to the through-port, is
\begin{align}
&{{R}_{CT}}\left( {{\omega }_{r}} \right)=\frac{{{P}_{C}}}{{{P}_{T}}} \nonumber \\
&=\frac{4{{\gamma }_{1}}{{\gamma }_{3}}}{{{\left| {{t}_{B}}\left( {{\omega }_{0}}-{{\omega }_{r}} \right)-j{{t}_{B}}\left( {{\gamma }_{v}}+{{\gamma }_{3}} \right)+S_e/t_B \right|}^{2}}}. \nonumber
\end{align}
Fig.~\ref{fig:s2} compares ${{R}_{LT}}$ and ${{R}_{CT}}$. Here we choose ${{r}_{L}}$=0.99 and ${{\omega }_{0}}/\left( 2{{\gamma }_{3}} \right)$=1$\times$10$^5$ so that the maximum of the Fano mirror reflectivity $\left| {{r}_{R}}\left( \omega  \right) \right|={{r}_{L}}$ (when $r_B$=0). As seen, both ${{R}_{LT}}$ and ${{R}_{CT}}$ are much larger than unity, and the cross-port, in general, exhibits the highest external quantum efficiency when working around the frequency point of the optimum linewidth ${{\omega }_{m}}$ (Fig.~\ref{fig:s2}(a)). When working around the peak of the Fano mirror reflectivity, the left mirror can give a higher external quantum efficiency (Fig.~\ref{fig:s2}(b)).

% The \nocite command causes all entries in a bibliography to be printed out
% whether or not they are actually referenced in the text. This is appropriate
% for the sample file to show the different styles of references, but authors
% most likely will not want to use it.
% \nocite{*}

\bibliographystyle{apsrev4-2}
\bibliography{LW_FL}% Produces the bibliography via BibTeX.

\end{document}